\documentclass[a4paper,12pt]{article}
\usepackage{hyperref}
\usepackage[a4paper,left=2.5cm,right=2.5cm,top=3cm,bottom=2.5cm]{geometry}
\usepackage{authblk}
\usepackage{amsmath,bm}
\usepackage{graphicx}
\usepackage{mathrsfs}
\usepackage{amssymb}
\usepackage{physics}
\usepackage{tensor}
\usepackage{xcolor}
\usepackage{tikz}
\usepackage{pgfplots}
\usepackage{amsthm}

\theoremstyle{remark}
\newtheorem*{remark}{\textbf{Remark}}

\title{Geometrically exact static isogeometric analysis of an arbitrarily curved spatial Bernoulli-Euler beam }

\author[1,2]{A.~Borković}
\author[1]{B.~Marussig}
\author[2,3]{G.~Radenković}

\affil[1]{Institute of Applied Mechanics, Graz University of Technology, Technikerstraße 4/II, 8010 Graz, Austria, aleksandar.borkovic@aggf.unibl.org, aborkovic@tugraz.at}
\affil[2]{University of Banja Luka, Faculty of Architecture, Civil Engineering and Geodesy, Department of Mechanics and Theory of Structures, 78000 Banja Luka, Bosnia and Herzegovina}
\affil[3]{Faculty of Civil Engineering, University of Belgrade, Bulevar kralja Aleksandra 73, 11000 Belgrade, Serbia}
\date{}                     
\begin{document}

	\newcommand{\ssub}[2]{{#1}_{#2}} 
	\newcommand{\vsub}[2]{\textbf{#1}_{#2}} 
	\newcommand{\ssup}[2]{{#1}^{#2}} 
	\newcommand{\vsup}[2]{\textbf{#1}^{#2}} 
	\newcommand{\ssupsub}[3]{{#1}^{#2}_{#3}} 
	\newcommand{\vsupsub}[3]{\textbf{#1}^{#2}_{#3}} 
	
	\newcommand{\veq}[1]{\bar{\textbf{#1}}} 
	\newcommand{\seq}[1]{\bar{#1}} 
	\newcommand{\ve}[1]{\textbf{#1}} 
	\newcommand{\sdef}[1]{#1^*} 
	\newcommand{\vdef}[1]{{\textbf{#1}}^*} 
	\newcommand{\vdefeq}[1]{{\bar{\textbf{#1}}}^*} 
	\newcommand{\trans}[1]{\textbf{#1}^\mathsf{T}} 
	\newcommand{\transmd}[1]{\dot{\textbf{#1}}^\mathsf{T}} 
	\newcommand{\mdvdef}[1]{\dot{\textbf{#1}}^*} 
	\newcommand{\mdsdef}[1]{\dot{#1}^*} 
	\newcommand{\mdv}[1]{\dot{\bm{#1}}} 
	\newcommand{\mdvni}[1]{\dot{\textbf{#1}}} 
	\newcommand{\mds}[1]{\dot{#1}} 
	
	\newcommand{\loc}[1]{\hat{#1}} 
	\newcommand{\md}[1]{\dot{#1}} 

	\newcommand{\ii}[3]{{#1}^{#2}_{#3}} 
	\newcommand{\iv}[3]{\textbf{#1}^{#2}_{#3}} 
	\newcommand{\idef}[3]{{#1}^{* #2}_{#3}} 
	\newcommand{\ivdef}[3]{\textbf{#1}^{* #2}_{#3}} 
	\newcommand{\iloc}[3]{\hat{#1}^{#2}_{#3}} 
	\newcommand{\ieq}[3]{\bar{#1}^{#2}_{#3}} 
	\newcommand{\ic}[3]{\tilde{#1}^{#2}_{#3}} 
	\newcommand{\icdef}[3]{\tilde{#1}^{* #2}_{#3}} 
	\newcommand{\iveq}[3]{\bar{\textbf{#1}}^{#2}_{#3}} 
	\newcommand{\ieqdef}[3]{\bar{#1}^{* #2}_{#3}} 
	\newcommand{\iveqdef}[3]{\bar{\textbf{#1}}^{* #2}_{#3}} 
	
	\newcommand{\ieqmddef}[3]{\dot{\bar{#1}}^{* #2}_{#3}} 
	\newcommand{\icmddef}[3]{\dot{\tilde{#1}}^{* #2}_{#3}} 
	\newcommand{\iveqmddef}[3]{\dot{\bar{\textbf{#1}}}^{* #2}_{#3}} 
	
	\newcommand{\ieqmd}[3]{\dot{\bar{#1}}^{#2}_{#3}} 
	\newcommand{\icmd}[3]{\dot{\tilde{#1}}^{#2}_{#3}} 
	\newcommand{\iveqmd}[3]{\dot{\bar{\textbf{#1}}}^{#2}_{#3}} 
	
	\newcommand{\imddef}[3]{\dot{#1}^{* #2}_{#3}} 
	\newcommand{\ivmddef}[3]{\dot{\textbf{#1}}^{* #2}_{#3}} 
	
	\newcommand{\imd}[3]{\dot{#1}^{#2}_{#3}} 
	\newcommand{\ivmd}[3]{\dot{\textbf{#1}}^{#2}_{#3}} 
	
	\newcommand{\iii}[5]{^{#2}_{#3}{#1}^{#4}_{#5}} 
	\newcommand{\iiv}[5]{^{#2}_{#3}{\textbf{#1}}^{#4}_{#5}} 
	\newcommand{\iivn}[5]{^{#2}_{#3}{\tilde{\textbf{#1}}}^{#4}_{#5}} 
	\newcommand{\iiieq}[5]{^{#2}_{#3}{\bar{#1}}^{#4}_{#5}} 
	\newcommand{\iiieqt}[5]{^{#2}_{#3}{\tilde{#1}}^{#4}_{#5}} 
	
	\newcommand{\eqqref}[1]{Eq.~\eqref{#1}} 
	\newcommand{\fref}[1]{Fig.~\ref{#1}} 

	\maketitle
	
\section*{Abstract}

The objective of this research is the development of a geometrically exact model for the analysis of arbitrarily curved spatial Bernoulli-Euler beams. The complete metric of the beam is utilized in order to include the effect of curviness on the nonlinear distribution of axial strain over the cross section. The exact constitutive relation between energetically conjugated pairs is employed, along with four reduced relations. The isogeometric approach, which allows smooth connections between finite elements, is used for the spatial discretization of the weak form. Two methods for updating the local basis are applied and discussed in the context of finite rotations. All the requirements of geometrically exact beam theory are satisfied, such as objectivity and path-independence. The accuracy of the formulation is verified by a thorough numerical analysis. The influence of the curviness on the structural response is scrutinized for two classic examples. If the exact response of the structure is sought, the curviness must be considered when choosing the appropriate beam model.

\textbf{Keywords}: spatial Bernoulli-Euler beam; strongly curved beams; geometrically exact analysis; analytical constitutive relation;

\section{Introduction}

Contemporary engineering is facing a growing demand for novel types of cost-effective structures which are simultaneously resistant and flexible. This trend is encouraged by the development of new structural materials, the constant improvement of existing ones, and novel discoveries related to structural form-finding. Curved spatial beams are crucial components of these structures and their application is fundamental in various fields of engineering, molecular physics, electronics, bio-mechanics, optics, etc. In order to apply beam models to these new challenges, more accurate mathematical and mechanical models will be required. 

First beam theories originate from the works of Euler and Bernoulli, later generalized by Kirchhoff, Clebsch, and Love \cite{1992dilla}. There is great nomenclature diversity for beam theories and a study that makes a careful attempt to classify these is given in \cite{2019meier}. Accordingly, the subject of the presented research is an arbitrarily curved and twisted beam with an anisotropic solid cross section, without warping. If the assumption of rigid cross sections is applied, Simo-Reissner (SR) theory is obtained. The addition of the constraint that the cross section remains perpendicular to the deformed axis results with the Bernoulli-Euler (BE) theory.

The nonlinear theory of shear deformable curved spatial beams is proposed by Reissner \cite{1981reissnera}. Simo completed his work and introduced the term geometrically exact beam theory which compromises a formulation for which \emph{the relationships between the configuration and the strain measures are consistent with the virtual work principle and the equilibrium equations at a deformed state regardless of the magnitude of displacements, rotations, and strains} \cite{1985simo}. This is often falsely referred to as a large strain theory despite the small strain assumption being present. Due to the presence of finite rotations, the configuration space of the GE beam theory is a special orthogonal (Lie) group, SO(3). It is a nonlinear, differentiable manifold, whose associated group elements are not additive or commutative, which lead to various types of algorithms, \cite{1986simo, 1988cardona, 1988simo,1988iura, 1995ibrahimbegovica, 1997ibrahimbegovic}. Recapitulation of these findings until 1997 is given in \cite{1997crisfield}.

It is argued in \cite{1999crisfielda, 1999jelenica} that all finite element (FE) implementations of the geometrically exact beam theory, existent at that time, are non-objective and path-dependent. An orthogonal interpolation scheme that is independent of the vector parameterization of a rotation manifold is suggested. This paved the way for the development of many alternative rotation interpolation strategies, \cite{2009ghosh, 2004romero, 2013zupan}. The corotational formulation employed in \cite{1999hsiao} and \cite{2014le} preserves objectivity by definition. Recently, an improvement to the corotational approach for shear deformable beams is given in \cite{2020magisano}. The carefully calculated corotational nodal rotations are interpolated which led to the satisfaction of all the requirements of the geometrically exact beam theory. The fact that many formulations do not deal with truly geometrically exact curved/twisted beams is discussed in \cite{2003kapaniaa}. 

The nonlinear analysis of spatial BE beams recently came under a similar focus of researchers. The problem of the invariance of classic curved beam theories is thoroughly elaborated in \cite{2012armerob} and \cite{2012armeroa}. Membrane locking is another problem of thin curved BE beam which is treated by various techniques such as assumed strain or stress fields and reduced/selective integration \cite{2012armerob}. Works \cite{2002weiss} and \cite{2011boyer} stand out due to their consistency with the geometrically exact approach. The most systematic up-to-date approach is the formulation by Meier et al. \cite{2014meier}. It satisfies many properties required by the geometrically exact analysis of a curved spatial slender BE beam. The effect of membrane locking is discussed in \cite{2015meiera} while the beam-to-beam contact problem using a torsion-free beam model is considered in \cite{2017meier}. A comprehensive review of beam theories is given by the same authors in \cite{2019meier}.

The isogeometric technique for the spatial discretization has attracted much attention \cite{2005hughes}. The nonlinear shear-deformable spatial beam is readily considered in the framework of isogeometric analysis (IGA) \cite{2016marino, 2017marino, 2017weeger, 2019marino, 2020vob, 2020tasora, 2020choi}. Regarding the nonlinear analysis of BE beams in the context of IGA, a torsion- and rotation-free spatial cable formulation is developed in \cite{2013raknesa} with attractive nonlinear dynamic applications. A spatial BE beam modeled as a ribbon with four degrees of freedom (DOF) is introduced in \cite{2013grecoa} and extended to the assemblies of beams in \cite{2014greco}. A mixed approach is considered in \cite{2016greco} while the consistent tangent operator is derived in \cite{2015grecoa}. Based on \cite{2013grecoa}, the authors in \cite{2016bauera} successfully performed nonlinear applications of a spatial curved BE beam. Recently, a multi-patch approach for the nonlinear analysis of BE beams is suggested \cite{2021vo}. The smallest rotation algorithm is used for an update of the basis. The nonlinear transformation between total cross-sectional rotations and unknown kinematics is defined in order to connect beams. An invariant geometric stiffness matrix is derived in \cite{2021yang} considering various end moments. However, the formulation is applied solely to the buckling analysis. The effect of initial curviness on the convergence properties of the solution procedure is considered in \cite{2021herath}.

These works represent the state of the art of curved spatial BE beam theories. In contrast to the presented research, most of the reviewed literature disregards the higher order terms with respect to the beam cross-sectional metric. The majority of them also do not consider truly geometric exact nonlinear analysis due to their lack of objectivity. Recently, Radenković and Borković contributed to the linear static analysis of arbitrarily curved BE beams \cite{2018radenkovicb, 2020radenkovicb}, based on foundations given in \cite{2017radenkovic}. The aim of this paper is to extend the developed formulation, which is applicable to strongly curved beams, to the nonlinear setting of geometrically exact beam theory. 

For curved beams, axial and bending actions are coupled due to the nonlinear distribution of strain along the cross section. This distribution depends on \textit{the curviness of beam} $Kd$. Here, $K$ is the curvature of beam axis and $d$ is the maximum dimension of the cross section in the planes parallel to the osculating plane \cite{2018radenkovicb}. Therefore, the term \textit{arbitrarily curved beam} actually refers to a beam with arbitrary curviness. With respect to this parameter, curved beams are readily classified as small-, medium- or big-curvature beams \cite{2010slivker}. We are here mostly concerned here with big-curvature, also known as \textit{strongly curved}, beams, for which $Kd > 0.1$. 

Although known for a long time, it is only recently that the strongly curved beams have been scrutinized in the framework of the modern numerical methods. Linear analysis of strongly curved plane beams is given in \cite{2016cazzania, 2018borkovicb, 2019borkovicb}, while nonlinear analysis is considered in \cite{2021borkovic}. The linear response of spatial beams is analyzed in \cite{2018radenkovicb}. Here, we will consider the nonlinear behavior of prismatic BE beams of which some are strongly curved, i.e., $Kd > 0.1$, either locally or globally. To the best of our knowledge, this is the first paper that deals with the geometrically exact analysis of strongly curved spatial BE beams within the framework of IGA. The paper is based on our previous works \cite{2018borkovicb, 2019borkovicb, 2018radenkovicb, 2017radenkovic}. Its main contribution is the extension of the foundations given in \cite{2017radenkovic, 2018radenkovicb} to the geometrically exact setting by defining the appropriate basis update and the consistent derivation of the stiffness matrix. The geometric stiffness is found by the careful variation of both the internal and the external virtual power with respect to the unknown metric. Special care is devoted to the variation of the twist variable and its gradient. The exact geometry of the spatial BE beam is utilized for the derivation of weak form of equilibrium which is solved by the Newton-Raphson and arc-length methods. A strict derivation of the constitutive relation allows us to derive reduced models and to assess the effects of these simplifications through the numerical analysis. Comparison with existing results confirms that the obtained formulation is reliable for the finite rotation analysis of arbitrarily curved beams. Moreover, the approach introduces an additional level of accuracy when dealing with strongly curved beams. The present formulation is geometrically exact in a sense that it strictly defines a relation between work conjugate pairs which allows analysis of arbitrarily large rotations and displacements \cite{1985simo}. Furthermore, by the careful implementation of the procedure for the basis update, the formulation is objective and path-independent \cite{2014meier}.

The paper is structured as follows. The next section presents the basic relations of the beam metric and this is followed by the description of the BE beam kinematics. The finite element formulation is given in Section 4 and numerical examples are presented in Section 5. The conclusions are presented in the last section.

\section{Metric of the beam continuum in the reference configuration}

The metric of a spatial beam model at some reference configuration, is elaborated. The position of the beam axis and the orientation of the cross sections are thus required. The definition of the position of a spatial curve is trivial, while the orientation of the cross sections can be defined in several ways. One approach is to define the orientation of the cross section at the beginning of the beam, and to choose an operator of some kind to describe its relative orientation along the beam \cite{2013grecoa}. An alternative approach is to define a reference basis at each point of the beam axis, and then to define the material basis with respect to the reference one \cite{2014meier}. 
The latter approach is applied here by using the Frennet-Serret frame as the reference frame \cite{2018radenkovicb}. This frame represents the intrinsic frame of the beam axis since one of its basis vectors, $\ve{n}$, is aligned with the curvature vector of the beam axis, $\iv{c}{}{} = K \iv{n}{}{}$.

Boldface lowercase and uppercase letters are used for vectors and tensors/matrices, respectively. An asterisk sign is used to designate deformed configurations. Greek index letters take values of 2 and 3, while Latin ones take values of 1, 2 and 3. Partial and covariant derivatives with respect to the convective coordinates are designated with $(\bullet )_{,m}$ and $(\bullet )_{\vert m}$, respectively. Material time derivative is marked as $\imd{(\bullet)}{}{}$.

For the details on the NURBS-based IGA modeling of curves, references \cite{2005hughes, 1995piegla} are recommended. 

\subsection{Metric of the beam axis}

\begin{figure}
	\includegraphics[width=\linewidth]{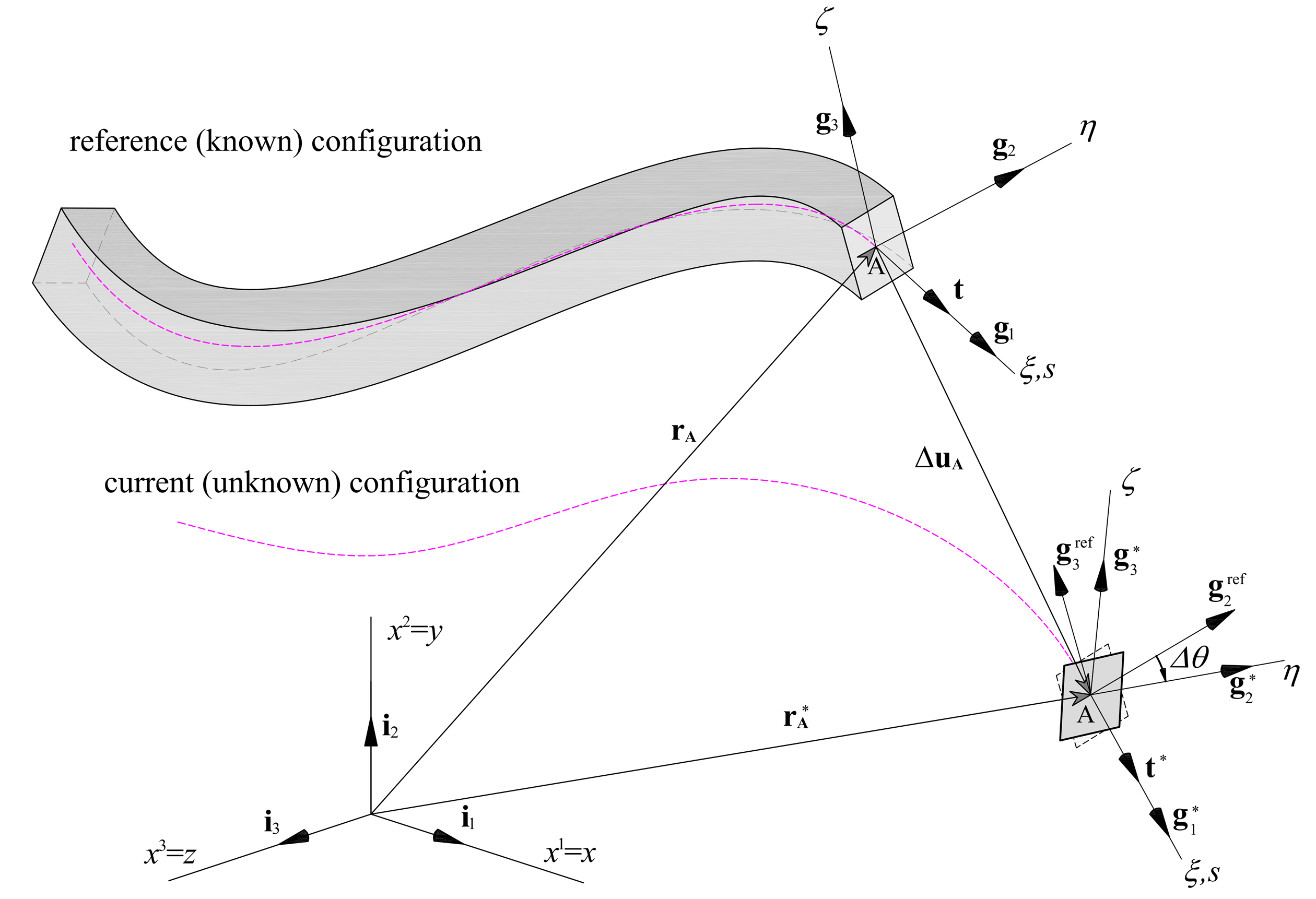}\centering
	\caption{Spatial beam, reference and current configurations.}
	\label{fig:begin}
\end{figure}

A beam axis is defined by its position vector $\iv{r}{}{} = \iv{r}{}{}(\xi) = \iv{r}{}{} (s)$, with $s$ being the arc-length coordinate while $\xi$ is some arbitrary parametric coordinate Fig.~\ref{fig:begin}. In general, setting up an arc-length parametrization of arbitrary curve is not possible in terms of elementary functions, hence $\xi$ is usually employed for the parametrization of the beam axis.

The position vector of the beam axis is given in global Cartesian coordinates by:
\begin{equation}
\label{eq: 1}
 \textbf{r} = \iv{r($\xi$)}{}{} = x^m (\xi) \textbf{i}_m = x^m \textbf{i}_m, \quad (x^1=x,\; x^2=y, \;x^3=z).
\end{equation}
Note that for the Cartesian coordinates, covariant and contravariant components are the same, and the position of indices is irrelevant. For every $C^1$ continuous curve, we can define a tangent vector:
\begin{equation}
\label{eq: 2}
\iv{g}{}{1} = \iv{r}{}{,1} = \frac{\dd{\iv{r}{}{}}}{\dd{\xi}} = \ii{x}{m}{,1} \iv{i}{}{m} =\frac{\dd{\iv{r}{}{}}}{\dd{s}} \frac{\dd{s}}{\dd{\xi}}=\frac{\dd{s}}{\dd{\xi}} \iv{t}{}{}.
\end{equation}
$\iv{g}{}{1}$ is the tangent vector of an arbitrary length, while $\iv{t}{}{}$ is the unit-length tangent of the beam axis. A function that defines the length of the basis vector $\iv{g}{}{1}$ is:
\begin{equation}
\label{eq: 3}
\norm{\iv{g}{}{1}} = \sqrt{\iv{g}{}{1} \cdot \iv{g}{}{1}} = \sqrt{\ii{g}{}{11}} = \sqrt{g} \implies \dd{s}=\sqrt{g} \dd{\xi}.
\end{equation}

At this point, let us introduce the well-known Frenet-Serret (FS) triad $\left(\iv{t}{}{}, \iv{n}{}{}, \iv{b}{}{} \right)$. Although not crucial for the present formulation, the intrinsic relation between the FS frame and the curvature of a line makes it useful for the presentation. The curvature vector of the beam axis is:
\begin{equation}
\label{eq:15}
\iv{c}{}{}=\frac{\dd{\iv{t}{}{}}}{\dd{s}}=\frac{\dd^2{\iv{r}{}{}}}{\dd{s^2}}=K\iv{n}{}{},
\end{equation}
where $\ve{n}$ is the unit-normal while $K$ is the curvature of axis. The third FS base vector, binormal, is now defined as:
\begin{equation}
\label{eq:16.1}
\iv{b}{}{}=\iv{t}{}{} \times \iv{n}{}{},
\end{equation}
and the well-known FS formulae follow:
\begin{equation}
\label{eq:18}
\begin{bmatrix}
\iv{t}{}{,s}\\
\iv{n}{}{,s}\\
\iv{b}{}{,s}
\end{bmatrix} = 
\begin{bmatrix}
0 & K & 0 \\
-K & 0 & \tau \\
0 & -\tau & 0
\end{bmatrix}
\begin{bmatrix}
\iv{t}{}{}\\
\iv{n}{}{}\\
\iv{b}{}{}
\end{bmatrix}
\Leftrightarrow \iv{t}{}{i,s} = \ii{K}{.j}{i} \iv{t}{}{j} = \ii{K}{}{ij} \iv{t}{}{j},
\end{equation}
where $\iv{t}{}{i} = \left(\iv{t}{}{},\iv{n}{}{},\iv{b}{}{}\right)$. Since the curvature tensor with respect to the arc-length parameter is skew-symmetric, we can define a pseudovector of curvature:
\begin{equation}
\label{eq:19}
\iv{k}{}{FS} = \ii{K}{i}{FS} \iv{t}{}{i} ,\quad \ii{K}{i}{FS} = \frac{1}{2} \ii{e}{ijl}{}\ii{K}{}{lj} ,\quad \iv{k}{}{FS} = \tau \iv{t}{}{} + K \iv{b}{}{},
\end{equation}
where $\tau$ is the torsion of the FS frame while $e$ is the permutation symbol. This allows us to rewrite the FS formulae as:
\begin{equation}
\label{eq:20}
\iv{t}{}{i,s} = \iv{k}{}{FS} \times \iv{t}{}{i}.
\end{equation}

In order to define the orientation of the cross section, we will introduce two unit base vectors, $\iv{g}{}{2}$ and $\iv{g}{}{3}$, which are aligned with the principal axes of inertia of the cross section. In this way, $\left( \iv{g}{}{1}, \iv{g}{}{2}, \iv{g}{}{3} \right)$ is a right-handed rectangular coordinate system at each point of a curve, see Fig.~\ref{fig:begin}. For curves with well-defined FS frame, we can specify an angle $\alpha$ between $\iv{t}{}{\beta}$ and $\iv{g}{}{\beta}$ vectors, and express the basis vectors as: 
\begin{equation}
\label{eq:14}
\begin{bmatrix}
\iv{g}{}{2}\\
\iv{g}{}{3}
\end{bmatrix} = 
\begin{bmatrix}
\cos \alpha  & \sin \alpha \\
-\sin \alpha & \cos \alpha
\end{bmatrix}
\begin{bmatrix}
\iv{n}{}{}\\
\iv{b}{}{}
\end{bmatrix}.
\end{equation}
With respect to the parametric coordinate, the metric tensor of a line and its reciprocal counterpart are:
\begin{equation}
\label{eq:4}
\ii{g}{}{ij}=
\begin{bmatrix}
\ii{g}{}{11} & 0 & 0\\
0 & 1 & 0 \\
0 & 0 & 1
\end{bmatrix} \implies 
\ii{g}{ij}{}=
\begin{bmatrix}
\ii{g}{11}{} & 0 & 0\\
0 & 1 & 0 \\
0 & 0 & 1
\end{bmatrix} ,
\quad \det(\ii{g}{}{ij}) = \ii{g}{}{11} = g, \quad \ii{g}{11}{} = \frac{1}{\ii{g}{}{11}}.
\end{equation}
The derivatives of the base vectors $\iv{g}{}{i}$ with respect to the parametric coordinate are:
\begin{equation}
\label{eq:def:christ}
\textbf{g}_{i,1} = x^k_{,i1} \textbf{i}_k =  \Gamma^{k}_{i1} \textbf{g}_k,
\end{equation}
where $\Gamma^{k}_{i1}$ are the Christoffel symbols of the second kind of a line. Since $\iv{g}{}{1}$ is not of unit-length, the resulting curvature tensor is not skew-symmetric:
\begin{equation}
\label{eq: def: derivatives of base vectors}
\begin{bmatrix}
\iv{g}{}{1,1}\\
\iv{g}{}{2,1}\\
\iv{g}{}{3,1}
\end{bmatrix}
=
\begin{bmatrix}
\iv{$\Gamma$}{1}{11} & \iv{$\Gamma$}{2}{11} & \iv{$\Gamma$}{3}{11}\\
\iv{$\Gamma$}{1}{21} & \iv{$\Gamma$}{2}{21} & \iv{$\Gamma$}{3}{21}\\
\iv{$\Gamma$}{1}{31} & \iv{$\Gamma$}{2}{31} & \iv{$\Gamma$}{3}{31}
\end{bmatrix}
\begin{bmatrix}
\iv{g}{}{1}\\
\iv{g}{}{2}\\
\iv{g}{}{3}
\end{bmatrix}
=
\begin{bmatrix}
\iv{$\Gamma$}{1}{11} & \ic{K}{}{3} & -\ic{K}{}{2}\\
-\ii{K}{}{3} & 0 & \ii{K}{}{1} \\
\ii{K}{}{2} & -\ii{K}{}{1} & 0
\end{bmatrix}
\begin{bmatrix}
\iv{g}{}{1}\\
\iv{g}{}{2}\\
\iv{g}{}{3}
\end{bmatrix}.
\end{equation}
Here, $\ii{K}{}{i}$ are the components of the curvature vector with respect to the material basis $\iv{k}{}{} = \ii{K}{i}{} \iv{g}{}{i} = \ii{K}{}{i} \iv{g}{i}{}$. Concretely, $\ii{K}{}{1}$ is the torsion of the material frame $\iv{g}{}{i}$, while $\ii{K}{}{\alpha}$ are components of curvature with respect to the $\iv{g}{}{\alpha}$ basis vectors. Furthermore, $\ic{K}{}{\alpha} = g \ii{K}{}{\alpha}$ are the same components but measured with respect to the parametric coordinate. The expressions for the components of curvature follow as:
\begin{equation}
\label{eq: 25}
\ii{K}{}{1} = \iv{g}{}{2,1}\cdot \iv{g}{}{3} , \quad \ic{K}{}{2} = -\iv{g}{}{1,1} \cdot \iv{g}{}{3}, \quad \ic{K}{}{3} = \iv{g}{}{1,1} \cdot \iv{g}{}{2}.
\end{equation}
It is now straightforward to find the relation between the components of the curvature vector with respect to the both coordinate systems \cite{2018radenkovicb}:
\begin{equation}
\label{eq:21}
\begin{aligned}
\ii{K}{1}{} &= \frac{1}{\sqrt{g}} \tau + \frac{1}{g} \ii{\alpha}{}{,1} \implies \ii{K}{}{1} = \ii{g}{}{11} \ii{K}{1}{}, \\
\ii{K}{2}{} &= \ii{K}{}{2} = K \sin \alpha, \quad \ii{K}{3}{} = \ii{K}{}{3} = K \cos \alpha.
\end{aligned}
\end{equation}
Finally, we can rewrite \eqqref{eq:20} as:
\begin{equation}
\label{eq:22}
\left(\frac{\iv{g}{}{i}}{\norm{\iv{g}{}{i}}}\right)_{,1} = \sqrt{g} \left( \iv{k}{}{} \times \frac{\iv{g}{}{i}}{\norm{\iv{g}{}{i}}} \right).
\end{equation}

\subsection{Metric of a generic point in beam continuum}

In order to completely define the geometry of a beam, we must define a coordinate system at each point of a beam continuum. Since the essence of structural theories is to reduce the problem, in this case from 3D to 1D, the complete metric of a beam should be defined by a set of reference quantities. It is common to use the metric of beam axis as the reference. For this, let us define an \textit{equidistant line} which is a set of points for which $ \left(\eta, \zeta \right) = const$. Its position and tangent base vectors are:
\begin{equation}
\label{eq:def:r_eq}
\begin{aligned}
\ve{r} \left(\xi,\eta,\zeta \right) &= \veq{r}\left(\xi\right) = \ve{r} + \eta \ve{g}_2 + \zeta \ve{g}_3, \\
\ve{g}_1 \left(\xi,\eta,\zeta \right) &= \veq{g}_1 = \iveq{r}{}{,1} = \ve{g}_1 + \eta \iv{g}{}{2,1} + \zeta \iv{g}{}{3,1} .
\end{aligned}
\end{equation}
By introducing the so-called \emph{initial curvature correction term} $g_0=1-\eta \ii{K}{}{3} + \zeta \ii{K}{}{2}$, the base vectors of an equidistant line are \cite{2003kapaniaa}: 
\begin{equation}
\label{eq:30}
\iveq{g}{}{1}=\ii{g}{}{0} \iv{g}{}{1} - \zeta \ii{K}{}{1} \iv{g}{}{2} + \eta \ii{K}{}{1} \iv{g}{}{3}, \quad \iveq{g}{}{2} = \iv{g}{}{2}, \quad \iveq{g}{}{3} = \iv{g}{}{3}.
\end{equation}
Evidently, these basis vectors are not orthogonal at a generic point, which complicates further reduction from 3D to 1D. In order to overcome this difficulty, a novel coordinate $\ii{\xi}{}{\lambda} = \ii{\xi}{}{\lambda} \left(\xi,\eta,\zeta\right)$ is utilized in \cite{2018radenkovicb}. It is a line orthogonal to the cross section at each point of a beam. By using the coordinate system $\left(\ii{\xi}{}{\lambda},\eta,\zeta\right)$, axial and torsional actions are decoupled. The tangent base vector of a $\ii{\xi}{}{\lambda}$ line is set to be equal to the tangent base vector of the beam axis $\iveq{g}{}{\lambda} = \iv{g}{}{1}$. The present derivations are based on this coordinate system.

\section{Bernoulli-Euler beam theory}

The classical BE assumption states that a cross section is rigid and remains perpendicular to the beam axis in the deformed configuration. This allows us to describe the complete beam kinematics by the translation of beam axis and the angle of twist of cross section.

\subsection{Kinematics of beam axis}

The metric of the deformed configuration is described in a similar manner to that of the reference one, by defining the triad $\left( \iv{g}{*}{1},\iv{g}{*}{2},\iv{g}{*}{3}\right)$ in the current configuration. This triad must be found by the update of some reference configuration, and this process is not uniquely defined. Regarding the tangent base vector, its definition is straightforward. The position and tangent base vectors of the beam axis in current configuration are:
\begin{equation}
\label{eq:def:r equidistant def1}
\begin{aligned}
\idef{\ve{r}}{}{} &=\idef{\ve{r}}{}{} (\xi) = \ve{r} + \iv{u}{}{},\\
\idef{\ve{g}}{}{1} &= \idef{\ve{g}}{}{1} (\xi) = \idef{\ve{r}}{}{,1} = \iv{g}{}{1} + \iv{u}{}{,1},
\end{aligned}
\end{equation}
where $\iv{u}{}{}$ is the displacement vector of the beam axis.

The other two basis vectors must be found by the rotation from some reference configuration:
\begin{equation}
\label{eq:40}
\ivdef{g}{}{\alpha} = \iv{R}{}{} \iv{g}{}{\alpha} , \quad \alpha=2,3, 
\end{equation}
where $\ve{R}$ is the rotation tensor or rotator. In general, this tensor belongs to the special orthogonal group SO(3) and several parameterization of it exists \cite{1995geradin, 1997ibrahimbegovic}. Since the SO(3) group is nonlinear, it is convenient to switch to some linear space.

Let us focus on the velocity field of the beam, since it is tangent to the displacement field. The velocity gradients along the $\eta$ and $\zeta$ directions are:
\begin{equation}
\label{eq:41}
\iv{v}{}{,\alpha} = \ivmddef{g}{}{\alpha} = \ivmd{R}{}{} \iv{g}{}{\alpha} + \iv{R}{}{} \ivmd{g}{}{\alpha} = \ivmd{R}{}{} \iv{g}{}{\alpha},
\end{equation}
while the velocity gradient along the $\xi$ coordinate follows as $\iv{v}{}{,1} = \ivmddef{g}{}{1} = \ivmd{u}{}{,1}$.
For each member of the SO(3) group, there exists an appropriate spinor $\bm{\Phi}$ which belongs to the so(3) group of skew-symmetric tensors \cite{2003kapaniaa}. This spinor represents the infinitesimal rotation and allows the exponential parameterization of the rotator:
\begin{equation}
\label{eq:415}
\iv{R}{}{} = e^{\bm{\Phi}} \implies \ivmd{R}{}{} =   \mdv{\Phi} \iv{R}{}{}.
\end{equation}
In this way, the velocity gradients \eqref{eq:41} become:
\begin{equation}
\label{eq:42}
\iv{v}{}{,\alpha} = \mdv{\Phi} \ivdef{g}{}{\alpha}.
\end{equation}
The material derivative of the spinor $\bm{\Phi}$ is the antisymmetric part of the velocity gradient - the angular velocity. Its components define the psudovector $\bm{\omega}$ with respect to the local triad, \cite{1985simo}:
\begin{equation}
\label{eq:43}
\imd{\Phi}{}{ij} = 
\begin{bmatrix}
0 & -\ii{\omega}{3}{} & \ii{\omega}{2}{} \\
\ii{\omega}{3}{} & 0 & -\ii{\omega}{1}{} \\
-\ii{\omega}{2}{} & \ii{\omega}{1}{} & 0
\end{bmatrix}, \quad
\bm{\omega}= \ii{\omega}{i}{} \ivdef{g}{}{i},
\end{equation}
where:
\begin{equation}
\label{eq:44}
\ii{\omega}{i}{} = \frac{1}{2} \ii{\epsilon}{ijk}{} \iloc{v}{}{k \vert j} \implies \ii{\omega}{1}{}= \frac{1}{\sqrt{g^*}} \iloc{v}{}{3 \vert 2}, \;\; \ii{\omega}{2}{}=-\frac{1}{\sqrt{g^*}} \iloc{v}{}{3 \vert 1}, \;\; \ii{\omega}{3}{} = \frac{1}{\sqrt{g^*}} \iloc{v}{}{2 \vert 1}.
\end{equation}
Here, $\epsilon$ is the Levi-Civita symbol while $\iloc{v}{}{k}$ are the components of velocity with respect to the local triad. It is evident from the last equation that two components of angular velocity, $\omega^2$ and $\omega^3$, depend on the velocity field of beam axis:
\begin{equation}
\label{eq:45}
\begin{aligned}
\ii{\omega}{2}{} &= -\frac{1}{\sqrt{g^*}} ~ \iloc{v}{}{3 \vert 1} = -\frac{1}{\sqrt{g^*}} ~\ivdef{g}{}{3} \cdot \iv{v}{}{,1},\\
\ii{\omega}{3}{} &= \frac{1}{\sqrt{g^*}} ~\iloc{v}{}{2 \vert 1} = \frac{1}{\sqrt{g^*}} ~\ivdef{g}{}{2} \cdot \iv{v}{}{,1}.
\end{aligned}
\end{equation}
This observation confirms that a rotation of a cross section of a BE beam belongs to the SO(2) group of in-plane rotations \cite{2014meier}. The 2D rotation occurs in the normal plane of the deformed beam axis which is uniquely defined with $\ivdef{g}{}{1}$. Therefore, the only independent component of the angular velocity for the BE beam is the one with respect to the tangent of the beam axis:
\begin{equation}
\label{eq:46}
\ii{\omega}{1}{} = \frac{1}{\sqrt{g^*}} ~ \iloc{v}{}{3 \vert 2} = \frac{1}{\sqrt{g^*}} ~ \omega.
\end{equation}
Besides the components of the velocity of the beam axis, this component of angular velocity represents the fourth DOF of the BE beam model. We will designate its physical counterpart simply with $\omega$. This quantity represents the angular velocity of a cross section with respect to the tangent of the beam axis. It is discussed in \cite{2018radenkovicb, 2020yang} that this quantity can be decomposed into two parts, one part being the angular velocity of the FS frame. Since this approach is not applicable for general spatial curves, we will consider $\omega$ as the complete twist of the cross section over the increment of time.

For the sake of further derivations, let us find the gradients of velocity with respect to the principal axes of inertia as functions of the generalized coordinates. By using Eqs. \eqref{eq:42} and \eqref{eq:45} we obtain:
\begin{equation}
\label{eq:461}
\begin{aligned}
\iv{v}{}{,2} &= \mdv{\Phi} \ivdef{g}{}{2} = \bm{\omega} \times \ivdef{g}{}{2} = - \frac{1}{g^*} \iloc{v}{}{2 \vert 1} \ivdef{g}{}{1} + \iloc{u}{}{3 \vert 2} \ivdef{g}{}{3} = - \frac{1}{g^*} \left(\ivdef{g}{}{2} \cdot \iv{v}{}{,1} \right) \ivdef{g}{}{1} + \omega \ivdef{g}{}{3}, \\
\iv{v}{}{,3} &= \mdv{\Phi} \ivdef{g}{}{3} = \bm{\omega} \times \ivdef{g}{}{3} = - \frac{1}{g^*} \iloc{v}{}{3 \vert 1} \ivdef{g}{}{1} - \iloc{u}{}{3 \vert 2} \ivdef{g}{}{2} = - \frac{1}{g^*} \left(\ivdef{g}{}{3} \cdot \iv{v}{}{,1} \right) \ivdef{g}{}{1} - \omega \ivdef{g}{}{2}.
\end{aligned}
\end{equation}
The gradients $\iv{v}{}{,21}$ and $\iv{v}{}{,31}$ are also required:
\begin{equation}
\label{eq: gradients v21 v31}
\begin{aligned}
\iv{v}{}{,21} &= \left[ -\frac{1}{g^*} \left( \ivdef{g}{}{2} \cdot \iv{v}{}{,1} \right) \ivdef{g}{}{1} + \ivdef{g}{}{3} \omega  \right]_{,1}\\
&= -\frac{1}{g^*} \bigg\{ \left( \idef{\Gamma}{1}{11} \ivdef{g}{}{1} +\icdef{K}{}{3} \ivdef{g}{}{2} -\icdef{K}{}{2} \ivdef{g}{}{3} \right) \left( \ivdef{g}{}{2} \cdot \iv{v}{}{,1} \right)  + \left[ \left( \idef{K}{}{1} \ivdef{g}{}{3} -\idef{K}{}{3} \ivdef{g}{}{1} \right) \cdot \iv{v}{}{,1} \right] \ivdef{g}{}{1} \\
& \;\;\;\; + \left( \ivdef{g}{}{2} \cdot \iv{v}{}{,11} \right) \ivdef{g}{}{1} -2 ~ \idef{\Gamma}{1}{11} \left( \ivdef{g}{}{2} \cdot \iv{v}{}{,1} \right) \ivdef{g}{}{1}  \bigg\} + \left( \idef{K}{}{2} \ivdef{g}{}{1} - \idef{K}{}{1} \ivdef{g}{}{2}\right) \omega + \ivdef{g}{}{3} \omega _{,1} ,\\
\iv{v}{}{,31} &= \left[ -\frac{1}{g^*} \left( \ivdef{g}{}{3} \cdot \iv{v}{}{,1} \right) \ivdef{g}{}{1} - \ivdef{g}{}{2} \omega  \right]_{,1} \\
&= -\frac{1}{g^*} \bigg\{  \left( \idef{\Gamma}{1}{11} \ivdef{g}{}{1} +\icdef{K}{}{3} \ivdef{g}{}{2} -\icdef{K}{}{2} \ivdef{g}{}{3} \right) \left( \ivdef{g}{}{3} \cdot \iv{v}{}{,1} \right)  + \left[ \left( \idef{K}{}{2} \ivdef{g}{}{1} - \idef{K}{}{1} \ivdef{g}{}{2}\right) \cdot \iv{v}{}{,1} \right] \ivdef{g}{}{1} \\
& \;\;\;\; + \left( \ivdef{g}{}{3} \cdot \iv{v}{}{,11} \right) \ivdef{g}{}{1} - 2~ \idef{\Gamma}{1}{11} \left( \ivdef{g}{}{3} \cdot \iv{v}{}{,1} \right)\ivdef{g}{}{1} \bigg\} + \left( \idef{K}{}{3} \ivdef{g}{}{1} - \idef{K}{}{1} \ivdef{g}{}{3}\right) \omega - \ivdef{g}{}{2} \omega _{,1} .
\end{aligned}
\end{equation}

\subsection{Basis update}

As discussed previously, the tangent base vector of the beam axis, $\iv{g}{*}{1}$, follows directly from the current position of the beam axis. The other two base vectors, $\ivdef{g}{}{\alpha}$, must be found by the rotation, \eqqref{eq:40}. Due to the fact that only one rotation is the generalized coordinate of BE beam, the parameterization of this rotation is significantly simplified in comparison with shear-deformable models. Namely, the base vectors $\ivdef{g}{}{\alpha}$ lie in the normal plane of the deformed beam axis and they are calculated as an in-plane rotation of some reference vectors $\iv{g}{ref}{\alpha}$. The definition of these reference vectors is not unique. Two procedures for the update of basis vectors are considered here, the Smallest Rotation (SR) and the Nodal Smallest Rotation Smallest Rotation Interpolation (NSRISR).  

\subsubsection{Smallest Rotation}

The SR mapping is commonly used for the modeling of BE beams \cite{2021vo}. The procedure consists of two steps. First, the triad from the previous configuration is rotated in order to align its tangent $\ve{t}$ with the tangent of the current configuration $\iv{t}{*}{}$. This is done in such a way that this rotation angle is minimized, and the reference vectors are:
\begin{equation}
\label{eq:47}
\iv{g}{ref}{\alpha} = \iv{g}{}{\alpha} -\frac{\ivdef{t}{}{} \cdot \iv{g}{}{\alpha}}{1+\ivdef{t}{}{} \cdot \iv{t}{}{}} \left(\iv{t}{}{} + \ivdef{t}{}{}\right)
\end{equation}
In the second step, this reference frame is rotated in its plane by the angle $\Delta \theta = \omega \Delta t$, where $\Delta t$ is the current time increment. As a result, the base vectors of the current configuration are found:
\begin{equation}
\label{eq:48}
\begin{bmatrix}
\ivdef{g}{}{2}\\
\ivdef{g}{}{3}
\end{bmatrix} = 
\begin{bmatrix}
\cos \Delta \theta  & \sin \Delta \theta \\
-\sin \Delta \theta & \cos \Delta \theta
\end{bmatrix}
\begin{bmatrix}
\iv{g}{ref}{2}\\
\iv{g}{ref}{3}
\end{bmatrix}.
\end{equation}
Note that this mapping has a singularity for $\ivdef{t}{}{} \cdot \iv{t}{}{}=-1$ which is of no practical interest if the reference configuration is appropriately defined \cite{2021vo}.

\subsubsection{Nodal Smallest Rotation Smallest Rotation Interpolation}

Unfortunately the SR procedure leads to the non-objectivity of the resulting formulation. Therefore, the NSRISR mapping is introduced in \cite{2014meier} in order to overcome the deficiency of the SR mapping. It is based on the considerations originally developed in \cite{1999crisfielda}. Concretely, although the continuum strain measures of the geometrically exact beam theory are objective, their finite element implementations are not. The problem is caused by the interpolation of the rotations between the current and some reference configurations. This approach, in general, includes rigid-body rotations. In order to solve this issue, the authors of \cite{1999crisfielda} suggested a linear interpolation of the relative rotation between the element nodes. Since this relative rotation is free from any rigid-body motion, the objectivity of the discretized strain measures is preserved.

The NSRISR algorithm consists of three steps. First, the SR procedure is applied to the $\iv{g}{}{\alpha}$ vectors at the start and the end of the finite element. The resulting vectors are designated as $\iv{g}{SR}{\alpha, start}$ and $\iv{g}{SR}{\alpha, end}$. Second, the SR mapping is applied to the vectors $\iv{g}{SR}{\alpha, start}$ to develop the reference frame $\iv{g}{ref}{\alpha} (\xi)$ at each point of the finite element, $\xi \in \left[\xi_{start},\xi_{end}\right] $. In order to update the basis correctly, this reference frame should be rotated by the angle $\Delta \theta + \Delta \theta_{c}$, where $\Delta \theta_{c}$ is the correction angle. This angle represents the difference between the reference frame $\iv{g}{ref}{\alpha} (\xi)$, and the one which follow from the classic SR procedure described in the previous Subsection. The correction angle $\Delta \theta_{c}$ is zero at the start of the element, while, at the end of an element, it can be simply calculated as the angle between the frames $\iv{g}{SR}{\alpha, end}$ and $\iv{g}{ref}{\alpha} (\xi_{end})$. Finally, the correction angle is linearly interpolated between known values at the start and the end, which leads to the function $\Delta \theta_{c} (\xi)$. In this way, the updated basis is calculated:
\begin{equation}
\label{eq:48a}
\begin{bmatrix}
\ivdef{g}{}{2}\\
\ivdef{g}{}{3}
\end{bmatrix} = 
\begin{bmatrix}
\cos \left(\Delta \theta + \Delta \theta_{c} \right)  & \sin \left(\Delta \theta + \Delta \theta_{c} \right) \\
-\sin \left(\Delta \theta + \Delta \theta_{c} \right) & \cos \left(\Delta \theta + \Delta \theta_{c} \right)
\end{bmatrix}
\begin{bmatrix}
\iv{g}{ref}{2}\\
\iv{g}{ref}{3}
\end{bmatrix}.
\end{equation}
Derivatives of the base vectors $\iv{g}{*}{\alpha}$ with respect to the parametric coordinate $\xi$ can be calculated straightforwardly from the last expression.

\subsection{Strain rates}

After the metric of the beam continuum is defined, the next step is to introduce a strain measure. Let us assign the convective property to the coordinates $\left(\xi, \eta, \zeta \right)$. For the convective coordinates, the Lagrange strain equals the difference between the current and reference metrics:
\begin{equation}
\label{eq:def:strain}
\ii{\epsilon}{}{ij} = \frac{1}{2} \left( \idef{g}{}{ij} - \ii{g}{}{ij} \right).
\end{equation}
At an arbitrary point of beam continuum, this strain is:
\begin{equation}
\label{eq:def:strain1}
\ieq{\epsilon}{}{ij} = \frac{1}{2} \left( \ieqdef{g}{}{ij} - \ieq{g}{}{ij} \right),
\end{equation}
while the components of the strain rate are equal to the material derivatives of strain:
\begin{equation}
\label{eq:def:strain2}
\ieq{d}{}{ij} = \ieqmd{\epsilon}{}{ij} = \frac{1}{2}  \ieqmddef{g}{}{ij}.
\end{equation}
Due to the BE hypothesis, the shear strain rates $\ii{d}{}{21}, \ii{d}{}{31}, \ii{d}{}{23}, \ii{d}{}{32}$ vanish.

The strains at a generic point of beam continuum are strictly derived in \cite{2018radenkovicb} with respect to the $\left( \ii{\xi}{}{\lambda} , \eta, \zeta \right)$ coordinates. We omit this derivation for the sake of brevity. The resulting  strain rates are:
\begin{equation}
\label{eq: 992}
\begin{aligned}
\ieq{d}{}{11} &= \frac{1}{\idef{g}{}{0}} \left[ \left( 1+\eta \idef{K}{}{3} - \zeta \idef{K}{}{2} \right) \ii{d}{}{11} - \eta \imd{\kappa}{}{3} + \zeta \imd{\kappa}{}{2}\right],\\
\ieq{d}{}{12} &= -\frac{1}{\idef{g}{}{0}} \zeta \imd{\kappa}{}{1}, \quad
\ieq{d}{}{13} = -\frac{1}{\idef{g}{}{0}} \eta \imd{\kappa}{}{1},
\end{aligned}
\end{equation}
where $\ii{d}{}{11}$ is the axial strain rate of beam axis, $\imd{\kappa}{}{1}$ is the rate of change of torsional curvature, and $\imd{\kappa}{}{\alpha}$ are the rates of changes of bending curvatures with respect to the principal axes of inertia. These four quantities represent \textit{the reference strain rates} of BE beam since they allow the calculation of the complete strain rate field of a beam.

The next step is to derive the relation between the reference strain rates and the generalized coordinates. For the axial strain rate, the derivation is simple:
\begin{equation}
\label{eq: 99}
\ii{d}{}{11} = \frac{1}{2} \imddef{g}{}{11} = \ivdef{g}{}{1} \cdot \iv{v}{}{,1},
\end{equation}
while the curvature components require more attention. By using \eqqref{eq: 25}, we obtain:
\begin{equation}
\label{eq: 991}
\begin{aligned}
\imd{\kappa}{}{1} &= \imddef{K}{}{1} = \iv{v}{}{,21} \cdot \ivdef{g}{}{3} + \ivdef{g}{}{2,1} \cdot \iv{v}{}{,3}, \\
\imd{\kappa}{}{2} &= \icmddef{K}{}{2} = -\iv{v}{}{,11} \cdot \ivdef{g}{}{3} - \ivdef{g}{}{1,1} \cdot \iv{v}{}{,3},  \\
\imd{\kappa}{}{3} &= \icmddef{K}{}{3} = \iv{v}{}{,11} \cdot \ivdef{g}{}{2} + \ivdef{g}{}{1,1} \cdot \iv{v}{}{,2},
\end{aligned}
\end{equation}
and after introducing Eqs.~\eqref{eq:461} and \eqref{eq: gradients v21 v31} into \eqqref{eq: 991}, the final expressions for the rates of curvature changes are:
\begin{equation}
\label{eq: rs1}
\begin{aligned}
\imd{\kappa}{}{1} &= \idef{K}{}{2} \left( \ivdef{g}{}{2} \cdot \iv{v}{}{,1} \right) + \idef{K}{}{3} \left( \ivdef{g}{}{3} \cdot \iv{v}{}{,1} \right) + \ii{\omega}{}{,1}, \\
\imd{\kappa}{}{2} &=  -\ivdef{g}{}{3} \cdot \left( \iv{v}{}{,11} - \idef{\Gamma}{1}{11} \iv{v}{}{,1}\right) + \icdef{K}{}{3} \omega,  \\
\imd{\kappa}{}{3} &= \ivdef{g}{}{2} \cdot \left( \iv{v}{}{,11} - \idef{\Gamma}{1}{11} \iv{v}{}{,1}\right) - \icdef{K}{}{2} \omega. 
\end{aligned}
\end{equation}

\subsection{Spatial discretization}

Using IGA, both geometry and velocity are here discretized with the same univariate NURBS functions $R_I$. However, different univariate NURBS functions $R^{\omega}_J$ are utilized for the approximation of the angular velocity:
\begin{equation}
\label{eq:def:u}
\begin{aligned}
\ve{r} &= \sum\limits_{I =1}^{N} R_{I} (\xi) \ve{r}_{I}, \quad \alpha = \sum\limits_{I =1}^{N} R_{I} (\xi) \alpha_{I},\\
\quad \ve{v} &= \sum\limits_{I =1}^{N} R_{I} (\xi) \ve{v}_{I}, \quad  \omega = \sum\limits_{J =1}^{M} \ii{R}{\omega}{J} (\xi) \omega_{J},
\end{aligned}
\end{equation}
where $\left(\bullet \right)_{I}$ stands for the value of quantity at $I^{th}$ control point. 
If we introduce a vector of generalized coordinates, $\iv{v}{\mathsf{T}}{\omega}= \left[\trans{v} \; \omega\right]$, and the matrix of basis functions $ \ve{N} $, the kinematic field of the beam can be represented as:
\begin{equation}
\iv{v}{}{\omega} = \ve{N} \ivmd{q}{}{},
\end{equation}
where:
\setcounter{MaxMatrixCols}{20}
\begin{equation}
\label{eq:def:u via matrices2}
\begin{aligned}
\trans{$\ivmd{q}{}{}$} &=
\begin{bmatrix}
\ivmd{q}{}{1} & \ivmd{q}{}{2} & ... & \ivmd{q}{}{I} & ... & \ivmd{q}{}{N} & \ii{\omega}{}{1} & \ii{\omega}{}{2} & ... & \ii{\omega}{}{J} & ... & \ii{\omega}{}{M}
\end{bmatrix}, \quad
\ivmd{q}{}{I}=
\begin{bmatrix}
\ii{v}{1}{I} & \ii{v}{2}{I} & \ii{v}{3}{I} 
\end{bmatrix} 
\\
\ve{N} &= 
\begin{bmatrix}
\iv{N}{}{1} & \iv{N}{}{2} & ... & \iv{N}{}{I} & ... & \iv{N}{}{N} & \iv{N}{\omega}{1} & \iv{N}{\omega}{2} & ... & \iv{N}{\omega}{J} & ... & \iv{N}{\omega}{M}
\end{bmatrix}, \\
\iv{N}{}{I} &=
\begin{bmatrix}
\iv{R}{}{I}  \\
\textbf{0}_{1\times3}  \\
\end{bmatrix}, \quad
\iv{R}{}{I} = 
\begin{bmatrix}
\ii{R}{}{I} & 0 & 0 \\
0 & \ii{R}{}{I} & 0\\
0 & 0 & \ii{R}{}{I}
\end{bmatrix}, \quad
\iv{N}{\omega}{J} = 
\begin{bmatrix}
\textbf{0}_{3\times1}  \\
\ii{R}{\omega}{J} 
\end{bmatrix}.
\end{aligned}
\end{equation}
It is important that the interpolation of the angular velocity $\omega$ must have $C^0$ interelement continuity in order for the NSRISR formulation to keep its objectivity. This fact is proved in \cite{2014meier} and it will be considered in Section 5 via the numerical analysis. The \emph{k}-refinement, a distinct feature of IGA, ensures the highest-possible continuity, $C^{p-1}$, while increasing the degree, $p$, of the spline space \cite{2005hughes}. By careful subsequent knot insertion, interelement continuity can be reduced at required points. Our FEM implementation employs this ability for the different discretization of the velocity and angular velocity fields over the same mesh. This is the reason why, in general, we allow that $N \ne M$, and $\ii{R}{}{I} \ne \ii{R}{\omega}{I}$ in \eqqref{eq:def:u}.

\section{Finite element formulation}

In line with the previous derivation, we will formulate isogeometric spatial BE element using the principle of virtual power. 

Let us start from the generalized Hooke law for the linear elastic material, also known as the Saint Venant-Kirchhoff material model:
\begin{equation}
\label{eq:stress strain relation}
\ieqmd{\sigma}{i}{j} = 2 \mu \ieq{d}{i}{j} + \lambda \ii{\delta}{i}{j} \ieq{d}{m}{m},
\end{equation}
where $\mu$ and $\lambda$ are  Lamé material parameters. By introducing the BE constraints $\ieqmd{\sigma}{2}{2}=\ieqmd{\sigma}{3}{3}=0$, the non-zero contravariant components of stress rate with respect to the $\left(  \ii{\xi}{}{\lambda}, \eta, \zeta \right)$ coordinates are \cite{2018radenkovicb}:
\begin{equation}
\label{eq:stress strain relation1}
\ieqmd{\sigma}{11}{} = E \ieq{g}{11}{} \ieq{g}{11}{} \ieq{d}{}{11}, \quad \ieqmd{\sigma}{12}{} = \mu \ieq{g}{11}{} \ieq{d}{}{12}, \quad  \ieqmd{\sigma}{13}{} = \mu \ieq{g}{11}{} \ieq{d}{}{13}.
\end{equation}

\noindent Here, $E$ is modulus of elasticity and $\mu$ corresponds to shear modulus.

\subsection{Principle of virtual power}

The principle of virtual power represents the weak form of the equilibrium. It states that at any instance of time, the total power of the external, internal and inertial forces is zero for any admissible virtual state of motion. If the inertial effects are neglected and body and surface loads are reduced to the beam axis, this can be written for a spatial BE beam as: 
\begin{equation}
\label{eq:virtual power}
\delta \ii{P}{}{} =  \int_{V}^{} \iv{\bm{$\sigma$}}{*}{} : \delta \iv{d}{}{} \dd{V} - \int_{\xi}^{} \left( \iv{p}{*}{} \cdot \delta \ve{v} + \iv{m}{*}{} \cdot \delta \bm{\omega} \right) \sqrt{g} \dd{\xi},
\end{equation}
where $\bm{\sigma} $ is the Cauchy stress tensor, $\ve{d}$ is the strain rate tensor, while $\ve{p}$ and $\ve{m}$ are the vectors of external distributed line forces and moments, respectively. All these quantities are defined at the current, unknown, configuration. By assuming that the load is deformation-independent, only the stress is linearized:
\begin{equation}
\label{eq:linearization of stress}
\iii{\bm{\sigma}}{}{}{*}{} \approx \iii{\bm{\sigma}}{}{}{}{} + 
\mdv{\sigma} \Delta \ii{t}{}{},  
\end{equation}
where $\mdv{\sigma}$ is the stress rate tensor which is calculated as the Lie derivative of current stress. Since the components of the stress rate tensor are equal to the material derivatives of the components of the stress tensor, \cite{2021radenkovicb}, the linearized form of the virtual power is:
\begin{equation}
\label{eq:linearized virtual power}
\begin{aligned}
& \int_{V}^{} \left( \ieqmd{\sigma}{11}{} \delta \ieq{d}{}{11} + \ieqmd{\sigma}{12}{} \delta \ieq{d}{}{12}  +  \ieqmd{\sigma}{13}{} \delta \ieq{d}{}{13} \right) \dd{V} \Delta t \\
&+ \int_{V}^{} \left( \ieq{\sigma}{11}{} \delta \ieq{d}{}{11} + \ieq{\sigma}{12}{} \delta \ieq{d}{}{12} + \ieq{\sigma}{13}{} \delta \ieq{d}{}{13} \right) \dd{V} = \int_{\xi}^{} \left( \iloc{p}{i}{} \delta \iloc{v}{}{i} + \iloc{m}{i}{} \delta \ii{\omega}{}{i} \right)\sqrt{g} \dd{\xi},
\end{aligned}
\end{equation}
where $\iloc{p}{i}{}$ and $\iloc{m}{i}{}$ are the components of distributed line forces and moments with respect to local coordinates. 

We will simplify the notation for the remainder of this paper, by neglecting the asterisks. We emphasize that this change in notation does not introduce any ambiguity since (i) the stress and strain rates are instantaneous quantities, while the known stress is calculated at the previous configuration, and (ii) all integrations are performed with respect to the metric of the current configuration, in accordance with the updated Lagrangian procedure \cite{2007bathea}.

By integrating the left-hand side of \eqqref{eq:linearized virtual power} with respect to the area of the cross section, the integrals over the 3D volume reduce to line integrals along the beam axis:
\begin{equation}
\label{eq:from 3D to 2D}
\begin{aligned}
& \int_{\xi}^{} \left( \icmd{N}{}{} \delta \ii{d}{}{11} + \icmd{M}{}{1} \delta \imd{\kappa}{}{1} + \icmd{M}{}{2} \delta \imd{\kappa}{}{2} + \icmd{M}{}{3} \delta \imd{\kappa}{}{3} \right) \sqrt{g} \dd{\xi} \Delta t \\
& + \int_{\xi}^{} \left( \ic{N}{}{} \delta \ii{d}{}{11} + \ic{M}{}{1} \delta \imd{\kappa}{}{1} + \ic{M}{}{2} \delta \imd{\kappa}{}{2} + \ic{M}{}{3} \delta \imd{\kappa}{}{3} \right) \sqrt{g}  \dd{\xi},
\end{aligned}
\end{equation}
where $\ic{N}{}{}$ and  $\ic{M}{}{j}$ are stress resultant and stress couples, that are energetically conjugated with the reference strain rates of the beam axis, $\ii{d}{}{11}$ and $\imd{\kappa}{}{j}$. $\icmd{N}{}{}$ and $\icmd{M}{}{j}$ are their respective rates given by: 
\begin{equation}
\label{eq: rates of section forces}
\begin{aligned}
\icmd{N}{}{} &= \int_{A}  \left( 1+\eta \ii{K}{}{3} - \zeta \ii{K}{}{2} \right) \ieqmd{\sigma}{11}{} \dd{\eta} \dd{\zeta}, \\
\icmd{M}{}{1} &= \int_{A} \left( \eta  \ieqmd{\sigma}{13}{} - \zeta \ieqmd{\sigma}{12}{} \right) \dd{\eta} \dd{\zeta}, \\ 
\icmd{M}{}{2} &=\int_{A}  \zeta  \ieqmd{\sigma}{11}{} \dd{\eta} \dd{\zeta}, \\ \icmd{M}{}{3} &=- \int_{A}  \eta  \ieqmd{\sigma}{11}{} \dd{\eta} \dd{\zeta}.
\end{aligned}
\end{equation}
If we introduce the following vectors: 
\begin{equation}
\label{eq:vectors of section forces and stran rates}
\begin{aligned}
\trans{f} &= 
\begin{bmatrix}
\ic{N}{}{} & \ic{M}{}{1} & \ic{M}{}{2} & \ic{M}{}{3}
\end{bmatrix}, \quad
\trans{e} = 
\begin{bmatrix}
\ii{d}{}{11} & \imd{\kappa}{}{1} & \imd{\kappa}{}{2} & \imd{\kappa}{}{3} 
\end{bmatrix}, \\
\trans{p} &= 
\begin{bmatrix}
\ii{p}{}{1} & \ii{p}{}{2} & \ii{p}{}{3} 
\end{bmatrix}, \quad
\trans{m} = 
\begin{bmatrix}
\ii{m}{}{1} & \ii{m}{}{2} & \ii{m}{}{3} 
\end{bmatrix},
\end{aligned}
\end{equation}	
\eqqref{eq:linearized virtual power} can be expressed in compact matrix form as:
\begin{equation}
\label{matrix form of linearized virtual power}
\int_{\xi}^{} \transmd{f} \delta \ve{e} \sqrt{g} \dd{\xi} \Delta t + \int_{\xi}^{} \trans{f} \delta \ve{e} \sqrt{g} \dd{\xi} = \int_{\xi}^{} \left( \trans{p}  \delta \ve{v} + \trans{m} \delta \bm{\omega} \right) \sqrt{g} \dd{\xi}. 
\end{equation}

\subsection{The relation between energetically conjugated pairs}

The geometrically exact relations \eqref{eq:from 3D to 2D} and \eqref{eq: rates of section forces} are crucial for the accurate formulation of structural beam theories. In particular, energetically conjugated pairs are defined rigorously and the appropriate constitutive matrix is guaranteed to be symmetric. By the introduction of Eqs.~\eqref{eq: 992} and \eqref{eq:stress strain relation1} into Eq. \eqref{eq: rates of section forces}, the exact relation between energetically conjugated pairs of stress and strain rates is obtained. The resulting symmetric constitutive matrix $\iv{D}{}{}$ is derived in \cite{2018radenkovicb}:
\begin{equation}
\label{eq: def: DA}
\ivmd{f}{}{} = \iv{D}{}{} \iv{e}{}{},  \: \:
\iv{D}{}{}= \frac{E}{g^2}
\begin{bmatrix}
A & 0 & \ii{I}{}{2} & -\ii{I}{}{3} \\
0 & \mu g \ii{I}{}{11}/E & 0 & 0 \\
\ii{I}{}{2} & 0 & \ii{I}{}{22} & -\ii{I}{}{23} \\
-\ii{I}{}{3} & 0 & -\ii{I}{}{23} & -\ii{I}{}{33} 
\end{bmatrix},
\end{equation}
where the geometric properties of cross section are the functions of curvature:
\begin{equation}
\label{eq: def: IA}
\begin{aligned}
A &=\int_{A}^{} \frac{\left( 1 + \eta \ii{K}{}{3} - \zeta \ii{K}{}{2} \right) ^2}{g_0} \dd{\eta} \dd{\zeta} , \quad
\ii{I}{}{2}=\int_{A}^{} \zeta \frac{ 1 + \eta \ii{K}{}{3} - \zeta \ii{K}{}{2} }{g_0} \dd{\eta} \dd{\zeta}, \\
\ii{I}{}{3} &= \int_{A}^{} \eta \frac{ 1 + \eta \ii{K}{}{3} - \zeta \ii{K}{}{2} }{g_0} \dd{\eta} \dd{\zeta}, \quad  \ii{I}{}{23} = \int_{A}^{}  \frac{ \eta \zeta }{g_0} \dd{\eta} \dd{\zeta}, \\
\ii{I}{}{22} &= \int_{A}^{}  \frac{ \zeta ^2 }{g_0} \dd{\eta} \dd{\zeta}, \quad \ii{I}{}{33} = \int_{A}^{}  \frac{ \eta ^2 }{g_0} \dd{\eta} \dd{\zeta}, \quad \ii{I}{}{11} = \int_{A}^{}  \frac{ \eta ^2 + \zeta ^2 }{g_0} \dd{\eta} \dd{\zeta}.
\end{aligned}
\end{equation}
If we introduce six integrals:
\begin{equation}
\label{eq: def: IAH}
\begin{aligned}
H_1 &=\int_{A}^{} \frac{1}{g_0} \dd{\eta} \dd{\zeta} , \quad
H_\eta=\int_{A}^{} \frac{ \eta }{g_0} \dd{\eta} \dd{\zeta}, \quad
H_{\eta\eta} =\int_{A}^{} \frac{\eta^2}{g_0} \dd{\eta} \dd{\zeta}, \\
H_\zeta &=\int_{A}^{} \frac{ \zeta }{g_0} \dd{\eta} \dd{\zeta} , \quad
H_{\zeta\zeta}=\int_{A}^{} \frac{ \zeta^2 }{g_0} \dd{\eta} \dd{\zeta}, \quad
H_{\eta\zeta}=\int_{A}^{} \frac{ \eta\zeta }{g_0} \dd{\eta} \dd{\zeta},
\end{aligned}
\end{equation}
the geometric properties in \eqqref{eq: def: IA} can be rewritten as:
\begin{equation}
\label{eq: def: IAHfinal}
\begin{aligned}
A &=H_1+2 K_3 H_\eta + \ii{K}{2}{3} \ii{H}{}{\eta\eta} - 2 K_2 \ii{H}{}{\zeta} + \ii{K}{2}{2} \ii{H}{}{\zeta\zeta} - 2 K_2 K_3 \ii{H}{}{\eta\zeta}  , \\
I_2 &= H_\zeta + K_3 \ii{H}{}{\eta\zeta} - \ii{K}{}{2} \ii{H}{}{\zeta\zeta}, \quad
I_3 =H_\eta + K_3 \ii{H}{}{\eta\eta} - \ii{K}{}{2} \ii{H}{}{\eta\zeta}, \\
\ii{I}{}{23} &= \ii{H}{}{\eta\zeta}, \quad \ii{I}{}{22} = \ii{H}{}{\zeta\zeta}, \quad \ii{I}{}{33} = \ii{H}{}{\eta\eta}, \quad \ii{I}{}{11} = \ii{H}{}{\zeta\zeta} + \ii{H}{}{\eta\eta}.
\end{aligned}
\end{equation}
It is emphasized that these integrals can be analytically computed for standard symmetric solid cross section shapes. The derived exact constitutive model is designated as $\ii{D}{a}{}$ further in the paper.

This strict derivation allows us to examine the influence of the exact constitutive relation. Four reduced constitutive models are considered for this purpose. The first and the simplest one is designated with $D^0$. This model is based on two assumptions: (i) $g_0 \rightarrow 1$, and (ii) the matrix \ve{D} is diagonal:
\begin{equation}
\label{eq: def: D0}
\begin{aligned}
A &=\int_{A}^{} \dd{\eta} \dd{\zeta} , \quad
\ii{I}{}{2}=\ii{I}{}{3}=\ii{I}{}{23}=0, \\
\ii{I}{}{22} &= \int_{A}^{}  \zeta ^2 \dd{\eta} \dd{\zeta}, \quad \ii{I}{}{33} = \int_{A}^{}  \eta ^2 \dd{\eta} \dd{\zeta}, \quad \ii{I}{}{11} = \int_{A}^{}  \left(\eta ^2 + \zeta ^2 \right) \dd{\eta} \dd{\zeta}.
\end{aligned}
\end{equation}
Therefore, this model completely disregards the coupling between bending and axial actions \cite{2014meier}. The second reduced model, $D^1$, also employs the first assumption of $D^0$, $g_0 \rightarrow 1$, but it keeps the coupling terms from the matrix $\ve{D}$. This $D^1$ model is readily utilized for the analysis of beams with small curvature \cite{2013grecoa, 2018radenkovicb}. The involved integrals simplify to:
\begin{equation}
\label{eq: def: D11}
\begin{aligned}
H_1 &= \int_{A}^{} \dd{\eta} \dd{\zeta} , \quad
H_\eta = \int_{A}^{} \eta \dd{\eta} \dd{\zeta} = 0, \quad
H_{\eta\eta} =\int_{A}^{} \eta^2 \dd{\eta} \dd{\zeta}, \\
H_\zeta &=\int_{A}^{} \zeta \dd{\eta} \dd{\zeta}=0, \quad
H_{\zeta\zeta}=\int_{A}^{} \zeta^2 \dd{\eta} \dd{\zeta}, \quad
H_{\eta\zeta}=\int_{A}^{} \eta \zeta \dd{\eta} \dd{\zeta}=0.
\end{aligned}
\end{equation}
Let us emphasize that these results are valid only when the integrals are calculated with respect to the principle axes of inertia. Additionally, the higher order terms with respect to the curvature in the property $A$, \eqqref{eq: def: IAHfinal}, are neglected for this constitutive model, which yields $A=H_1$. 

In order to assess the effect of the higher order terms, we introduce two additional reduced models, $D^2$ and $D^3$, which employ a Taylor approximation of the exact expressions \eqref{eq: def: IAHfinal}. To be precise, the model $D^2$ is based on the $1^{st}$ order Taylor approximation of integrands which results in:
\begin{equation}
\label{eq: def: D21}
\begin{aligned}
H_1 &= \int_{A}^{}  \dd{\eta} \dd{\zeta} , \quad
H_\eta =\int_{A}^{} \left(  \ii{K}{}{3} \eta^2 \right) \dd{\eta} \dd{\zeta}, \quad
H_{\eta\eta} = \int_{A}^{}  \eta^2  \dd{\eta} \dd{\zeta}, \\
H_{\zeta} &= \int_{A}^{} \left( -\ii{K}{}{2} \zeta^2 \eta^2 \zeta^2 \right) \dd{\eta} \dd{\zeta} ,\quad
H_{\zeta\zeta}=\int_{A}^{}  \zeta^2   \dd{\eta} \dd{\zeta}, \quad
H_{\eta\zeta}= \int_{A}^{}  \eta \zeta  \dd{\eta} \dd{\zeta} = 0.
\end{aligned}
\end{equation}
The model $D^3$, on the other hand, follows from the $2^{nd}$ order Taylor approximation of the integrands:
\begin{equation}
\label{eq: def: D31}
\begin{aligned}
H_1 &= \int_{A}^{} \left(  1+ \ii{K}{2}{3} \eta^2 + \ii{K}{2}{2} \zeta^2 + 6  \ii{K}{2}{2} \ii{K}{2}{3} \eta^2 \zeta^2\right)  \dd{\eta} \dd{\zeta} , \\
H_\eta &=\int_{A}^{} \left(  \ii{K}{}{3} \eta^2  + 3 \ii{K}{2}{2} \ii{K}{}{3} \eta^2 \zeta^2 \right) \dd{\eta} \dd{\zeta}, \\
H_{\eta\eta} &= \int_{A}^{} \left( \eta^2 + \ii{K}{2}{2}  \eta^2 \zeta^2   \right) \dd{\eta} \dd{\zeta}, \\
H_{\zeta} &= \int_{A}^{} \left( -\ii{K}{}{2} \zeta^2 - 3 \ii{K}{}{2} \ii{K}{2}{3} \eta^2 \zeta^2 \right) \dd{\eta} \dd{\zeta} , \\
H_{\zeta\zeta}&=\int_{A}^{} \left( \zeta^2 + \ii{K}{2}{3} \eta^2 \zeta^2 \right) \dd{\eta} \dd{\zeta}, \\
H_{\eta\zeta}&=\int_{A}^{} \left( -2\ii{K}{}{2} \ii{K}{}{3} \eta^2 \zeta^2 \right) \dd{\eta} \dd{\zeta}.
\end{aligned}
\end{equation}

For beams with small curviness and circular cross section, the term $I_{11}$ reduces to the polar moment of area $I_0$, \eqqref{eq: def: D0}. However, for all the other shapes of cross section, this fact does not hold. Instead, the term $I_{11}$ is usually replaced with the so-called \emph{torsional constant}, $J$, which must be calculated approximately \cite{2021vo}. In the presented research, $J$ is used for the models $D^0$ and $D^1$. For the other three constitutive models, we have scaled the geometric property $I_{11}$ with the ratio $J/I_0$, in order to keep the influence of the curviness. However, this aspect did not affect any of our numerical experiments, even for strongly curved beams. Hence, these studies are skipped in the numerical experiments section, for the sake of brevity.

\subsection{Discrete equations of motion}

Let us turn now to the spatially discretized setting. First, we will define the matrix $\iv{B}{}{L}$ which relates the reference strain rates with the generalized coordinates at the control points by using Eqs.~\eqref{eq: 99}, \eqref{eq: rs1} and \eqref{eq:def:u via matrices2}:
\begin{equation}
\label{eq: vector of reference strains matrix form}
\ve{e} = \iv{B}{}{L} \ivmd{q}{}{}= \ve{H} \ve{B} \ivmd{q}{}{} ,
\end{equation}
where:
\setcounter{MaxMatrixCols}{20}
\begin{equation}
\label{eq: vector of reference strains matrix form2}
\begin{aligned}
\ve{B} &= 
\begin{bmatrix}
\iv{B}{}{1} & \iv{B}{}{2} & ... & \iv{B}{}{I} & ... & \iv{B}{}{N} & \iv{B}{\omega}{1} & \iv{B}{\omega}{2} & ... & \iv{B}{\omega}{J} & ... & \iv{B}{\omega}{M}
\end{bmatrix}, \\
\iv{B}{}{I} &=
\begin{bmatrix}
\iv{R}{}{I,1}   \\
\iv{R}{}{I,11}  \\
\textbf{0}_{2\times3}  \\
\end{bmatrix}, \quad
\iv{B}{\omega}{J} = 
\begin{bmatrix}
\textbf{0}_{6\times1}  \\
\ii{R}{\omega}{J} \\
\ii{R}{\omega}{J,1}  \\
\end{bmatrix}, \quad
\ve{H} =
\begin{bmatrix}
\trans{$\iv{g}{}{1}$} & \textbf{0}_{1\times3} & 0 & 0 \\
\ii{K}{}{2} \trans{$\iv{g}{}{2}$} + \ii{K}{}{3} \trans{$\iv{g}{}{3}$} & \textbf{0}_{1\times3} &  0 & 1 \\
\ii{\Gamma}{1}{11} \trans{$\iv{g}{}{3}$} &  -\trans{$\iv{g}{}{3}$} & \ic{K}{}{3}  &  0\\
-\ii{\Gamma}{1}{11} \trans{$\iv{g}{}{2}$} &  \trans{$\iv{g}{}{2}$} & -\ic{K}{}{2}  &  0
\end{bmatrix}.
\end{aligned}
\end{equation}
Since the strain rate is a function of the generalized coordinates as well as the metric, we must vary it with respect to both arguments. By the variation with respect to the generalized coordinates, a linear (material) part of the tangent stiffness is obtained. Variation with respect to the geometry results with the geometric stiffness matrix \cite{2021radenkovicb}. 

Clearly, variation of the reference strain rates in Eqs.~\eqref{eq: 99} and \eqref{eq: rs1} with respect to the generalized coordinates is trivial. On the other hand, variation with respect to the metric is more involved and it is given in detail in Appendix A. Formally, the variation of strain rate can be written as:
\begin{equation}
\label{eq: 1 e=Hw, BL=HB}
\delta \ve{e} = \delta \left( \iv{B}{}{L} \ivmd{q}{}{} \right) = \delta \ve{H} \ve{B} \ivmd{q}{}{} + \ve{H} \ve{B} \delta \ivmd{q}{}{}.
\end{equation}
Let us define the matrix of basis functions $\ve{B}_G$, which can be obtained by removing the $8^{th}$ row from the matrix $\ve{B}$ in \eqqref{eq: vector of reference strains matrix form2}, and the matrix of generalized section forces $\ve{G}$ with elements $\ii{G}{}{ij}$, cf. Appendix A. Now, by the insertion of \eqqref{eq: 1 e=Hw, BL=HB} into \eqqref{matrix form of linearized virtual power}, we can reformulate the term generated by the known stress and the variation of the strain rate with respect to the metric:
\begin{equation}
\label{eq: part of vp generated by known stress and variation of strain rate}
\int_{\xi}^{} \trans{f} \delta \iv{H}{}{} \ve{B} \ivmd{q}{}{} \sqrt{g} \dd{\xi} = \transmd{q} \int_{\xi}^{} \trans{$\iv{B}{}{G}$} \iv{G}{}{} \iv{B}{}{G} \sqrt{g} \dd{\xi} \delta \ivmd{q}{}{} \Delta t =
\transmd{q} \iv{K}{}{G} \delta \ivmd{q}{}{}  \Delta t,
\end{equation}
where $\iv{K}{}{G}$ is the geometric stiffness matrix. The complete derivation of this term is given in Appendix A.

Now, the integrands in the equation of the virtual power \eqref{matrix form of linearized virtual power} reduce to:
\begin{equation}
\label{eq: terms of VP rewritten}
\begin{aligned}
\transmd{f} \delta \ve{e} &\approx  \transmd{q} \trans{$\iv{B}{}{L}$} \ve{D} \iv{B}{}{L} \delta \ivmd{q}{}{},\\
\trans{f} \delta \ve{e}  &=  \trans{f} \left( \delta \iv{B}{}{L} \ivmd{q}{}{} + \iv{B}{}{L} \delta \ivmd{q}{}{}\right)  =  \transmd{q} \trans{$\iv{B}{}{G}$} \ve{G} \iv{B}{}{G} \delta \ivmd{q}{}{} \Delta t + \trans{f} \iv{B}{}{L} \delta \ivmd{q}{}{} ,
\end{aligned}
\end{equation}
where the first term is linearized by neglecting the variation of strain rate with respect to the metric.

Regarding the external virtual power, if we vary it with respect to the kinematics, the vector of the external load, $\ve{Q}$, is recovered:
\begin{equation}
\label{eq: virtual equilibrium0}
\int_{\xi}^{} \left( \trans{p}  \delta \ve{v} + \trans{m} \delta \bm{\omega} \right) \sqrt{g} \dd{\xi} = \trans{\ve{Q}} \delta \ivmd{q}{}{},
\end{equation}
and it is the same as in the linear analysis \cite{2018radenkovicb}. However, if we vary the external virtual power with respect to the geometry, a contribution to the geometric stiffness is obtained. The derivation of this term is given in Appendix B. Its implementation in the formulation is important for the successful convergence of the nonlinear solver.

Finally, the discretized and linearized equation of equilibrium is:
\begin{equation}
\label{eq: virtual equilibrium}
\transmd{q} \int_{\xi}^{} \left( \trans{$\iv{B}{}{L}$} \ve{D} \iv{B}{}{L} + \trans{$\iv{B}{}{G}$} \ve{G} \iv{B}{}{G} \right) \sqrt{g} \dd{\xi} \delta \ivmd{q}{}{} \Delta t =\trans{\ve{Q}} \delta \ivmd{q}{}{} - \int_{\xi}^{} \trans{f} \iv{B}{}{L} \sqrt{g} \dd{\xi} \delta \ivmd{q}{}{},
\end{equation}
which can be readily written in the standard form:
\begin{equation}
\label{eq:standard form of equlibrium}
\iv{K}{}{T} \Delta \ve{q} = \ve{Q} - \ve{F}, \quad \left( \Delta \ve{q} = \ivmd{q}{}{} \Delta t\right),
\end{equation}
where:
\begin{equation}
\label{eq: Kt}
\iv{K}{}{T} = \int_{\xi}^{} \trans{$\iv{B}{}{L}$} \ve{D} \iv{B}{}{L} \sqrt{g} \dd{\xi} + \int_{\xi}^{} \trans{$\iv{B}{}{G}$} \ve{G} \iv{B}{}{G} \sqrt{g} \dd{\xi}, 
\end{equation}
is the tangent stiffness matrix and:
\begin{equation}
\label{eq: Q and F}
\ve{F} = \int_{\xi}^{} \trans{$\iv{B}{}{L}$} \ve{f} \sqrt{g} \dd{\xi},
\end{equation}
is the vector of internal forces. The vector $\Delta \ve{q}$ in \eqqref{eq:standard form of equlibrium} contains increments of displacements and twist at control points with respect to the reference configuration. Due to the approximation introduced with \eqqref{eq: terms of VP rewritten}, the solution of \eqqref{eq:standard form of equlibrium} does not satisfy the principle of virtual power and some numerical procedure must be applied. Here, both the Newton-Raphson and arc-length methods are employed.

\section{Numerical examples}

The aim of the following numerical studies is to validate the presented approach and to examine the influence of the curviness on the structural response. The boundary conditions are imposed in a well-known manner and the rotations are treated with special care since two components are not utilized as DOFs \cite{2018radenkovicb}. Components of external moments with respect to the principal axes of inertia are applied as force couples which must be updated at each iteration \cite{2021borkovic}. Standard Gauss quadrature with $p+1$ integration points per element are used here.

All the results are presented with respect to the load proportionality factor ($LPF$), rather than to the load intensity itself. Since the time is a fictitious quantity in the present static analysis, strain and stress rates are equal to strains and stresses, respectively. Besides, increments of displacement and twist are equal to velocity and angular velocity.

Two approaches are considered here in the context of the reference configuration for the basis update, the total and the incremental. For the total approach, the reference configuration is the initial unstressed configuration \cite{2002ritto-correa}. For the incremental approach, the previously converged configuration is adopted as the reference one \cite{1988cardona}. The total approach should guarantee that the results are path-independent, since no history of deformation is included in the reference configuration. 

Furthermore, four element formulations are considered: $SR - C^{p-1}$, $SR - C^0$, $NSRISR - C^{p-1}$, and $NSRISR - C^0$.
These formulations vary in the basis update algorithm ($SR$ or $NSRISR$), and on the interelement continuity used for twist ($C^0$ or $C^{p-1}$). For the displacement of the beam axis, the highest available interelement continuity is applied exclusively. If not stated otherwise, the order of the spline functions used for the approximation of the twist is the same as that used for the displacement of the beam axis.

In some examples, the error of the position vector $\iv{r}{}{}$ is calculated using the relative $L^2$-error norm, as proposed in \cite{2014meier}:
\begin{equation}
\label{eq: l2 definition}
\ii{\norm{e}}{2}{} =\frac{1}{u_{max}} \sqrt{\frac{ \int_{0}^{L} \norm{\iv{r}{}{h}-\iv{r}{}{ref}}^2 \dd{s}}{L}},
\end{equation}
where $L$ is the length of the beam and $u_{max}$ is the maximum value of the displacement component. $\iv{r}{}{h}$ represents the approximate solution for the position of the beam axis while $\iv{r}{}{ref}$ is the appropriate reference solution.

\subsection{Objectivity}

The objectivity of the formulation can be considered as the invariance with respect to the rigid-body motions. Thus, if the beam is subjected to a rigid-body motion, no strain should appear. Invariance with respect to the translations is readily satisfied while the invariance with respect to the rotation requires special attention \cite{2012armeroa}. 

\begin{figure}
	\includegraphics[width=\linewidth]{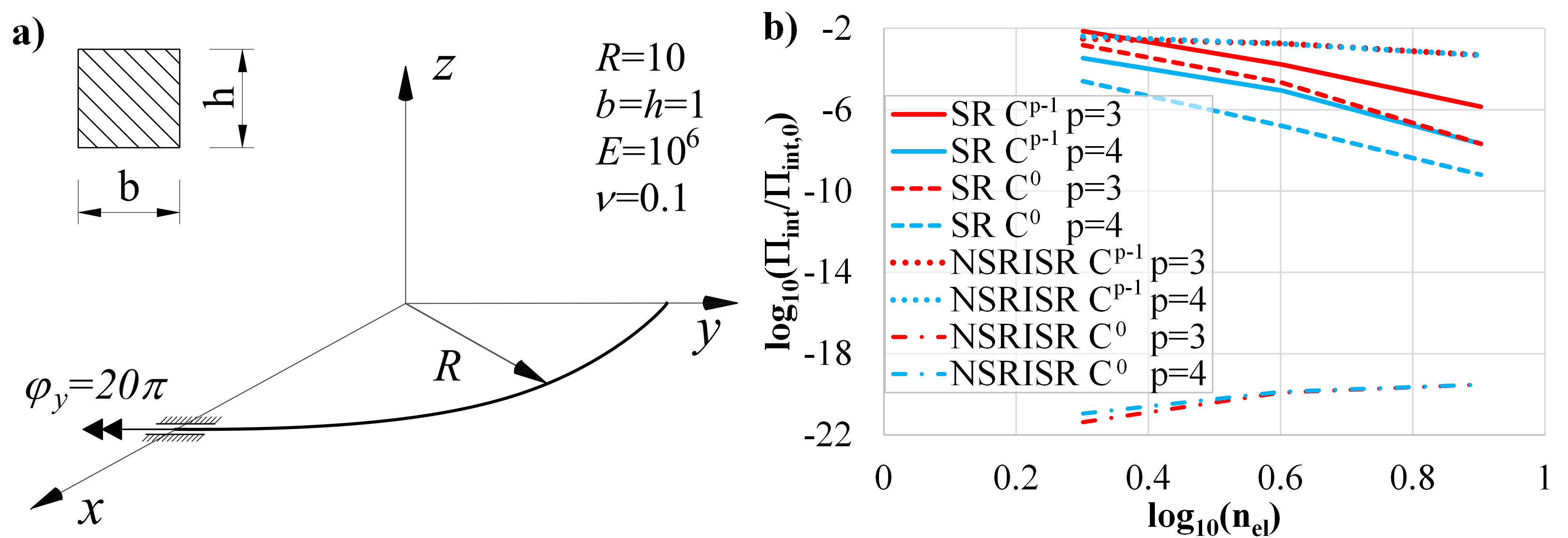}\centering
	\caption{Objectivity. a) Geometry and material properties. b) Ratio of internal strain energy for the different formulations with respect to the number of elements.}
	\label{fig:objectivity:disposition}
\end{figure}

The present example was introduced in \cite{2014meier}. A quarter-circular cantilever beam is rotated ten times around its clamped end with respect to the $y$-direction, see Fig.~\ref{fig:objectivity:disposition}a.
For an objective formulation, there should be no internal strain energy $\ii{\Pi}{}{int}$. 
Due to the extremely large rotations involved in this example, only the incremental update of rotations is considered. The non-homogeneous boundary condition, $\varphi_y=20\pi$, is applied in 100 increments. Two NURBS orders are considered, $p=3$ and $p=4$. 

The internal strain energy in the final configuration is plotted in Fig.~\ref{fig:objectivity:disposition}b with respect to the number of elements. The obtained results are scaled with the value $\ii{\Pi}{}{int,0}=EI\pi \slash (4R)$. This is the internal strain energy of an initially straight beam that is bent into a quarter circle with tip moment \cite{2014meier}. These results confirm that the $NSRISR-C^0$ formulation is indeed objective and the internal strain energy equals zero up to the machine precision. The beam deforms in all the other formulations considered. Furthermore, the results indicate that the problem with the representation of rigid rotations is mitigated for all formulations when the number of elements is increased. This is a well-known fact, which sometimes justifies the application of non-objective formulations in quasi-static analyses \cite{2020erdelj}.

\begin{figure}
	\includegraphics[width=\linewidth]{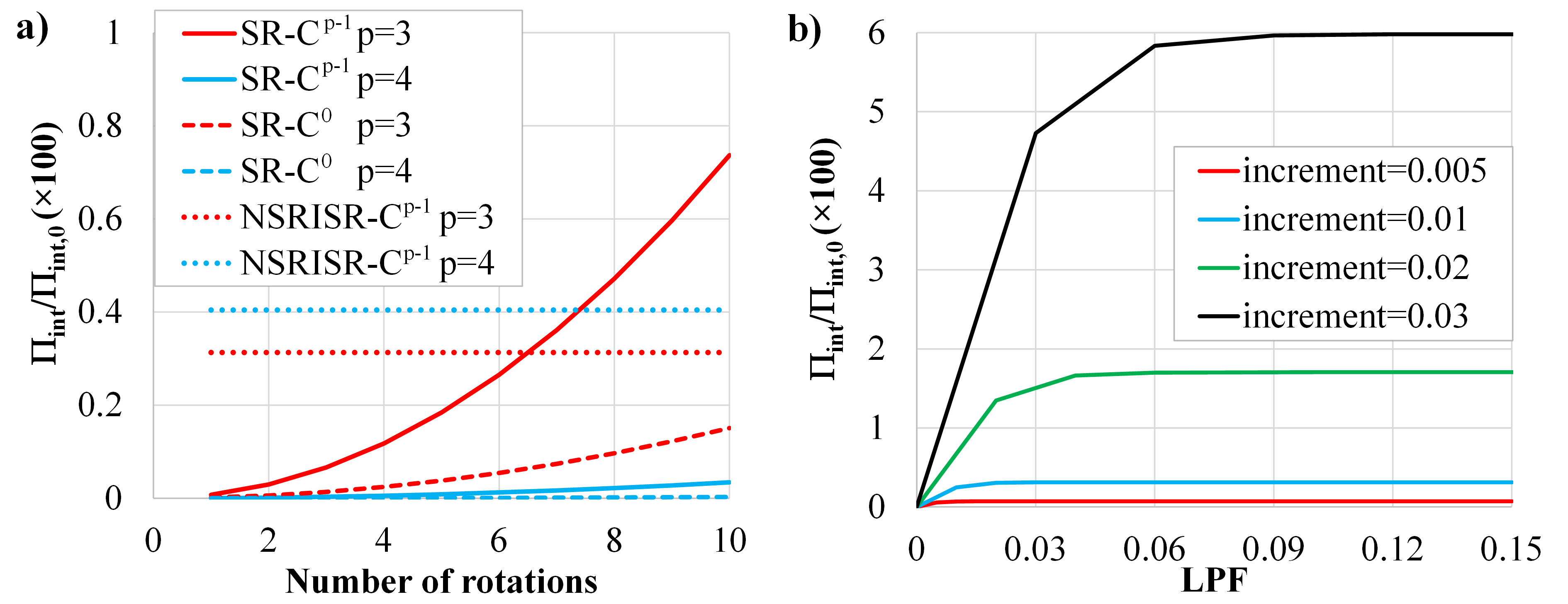}\centering
	\caption{Objectivity. a) Ratio of internal strain energy generated in model for the different formulations with respect to the number of rotation. b) Ratio of internal strain energy for $NSRISR-C^{p-1}$ formulation using two cubic elements for different increment sizes.}
	\label{fig:objectivity 2}
\end{figure}

Fig.~\ref{fig:objectivity 2}a shows the evolution of the internal strain energy with the number of rotations for different formulations and NURBS orders. 
Note that the $SR-C^{p-1}$ formulation with cubic NURBS elements generates around 0.7 \% of the reference strain energy, while it is less than 0.2 \% for $SR-C^{0}$ case. A similar behavior can be observed for quartic elements. It follows that the $SR$ formulation greatly benefits from the reduced interelement continuity of the twist variable.

The behavior of the $NSRISR-C^{p-1}$ formulation is also worth mentioning. The results suggest that the initially accumulated internal strain energy, during first few increments, remains constant. In order to examine this more thoroughly, the strain energies for the $NSRISR-C^{p-1}$ formulation and four different increment sizes are plotted in Fig.~\ref{fig:objectivity 2}b for $LPF<0.15$. It is apparent that the error increases with the size of the increment while asymptotically approaching constant value. For this formulation, the beam deforms at the beginning of the loading process and rotates afterwards without the additional straining.

\subsection{Path-independence}

\begin{figure}
	\includegraphics[width=7cm]{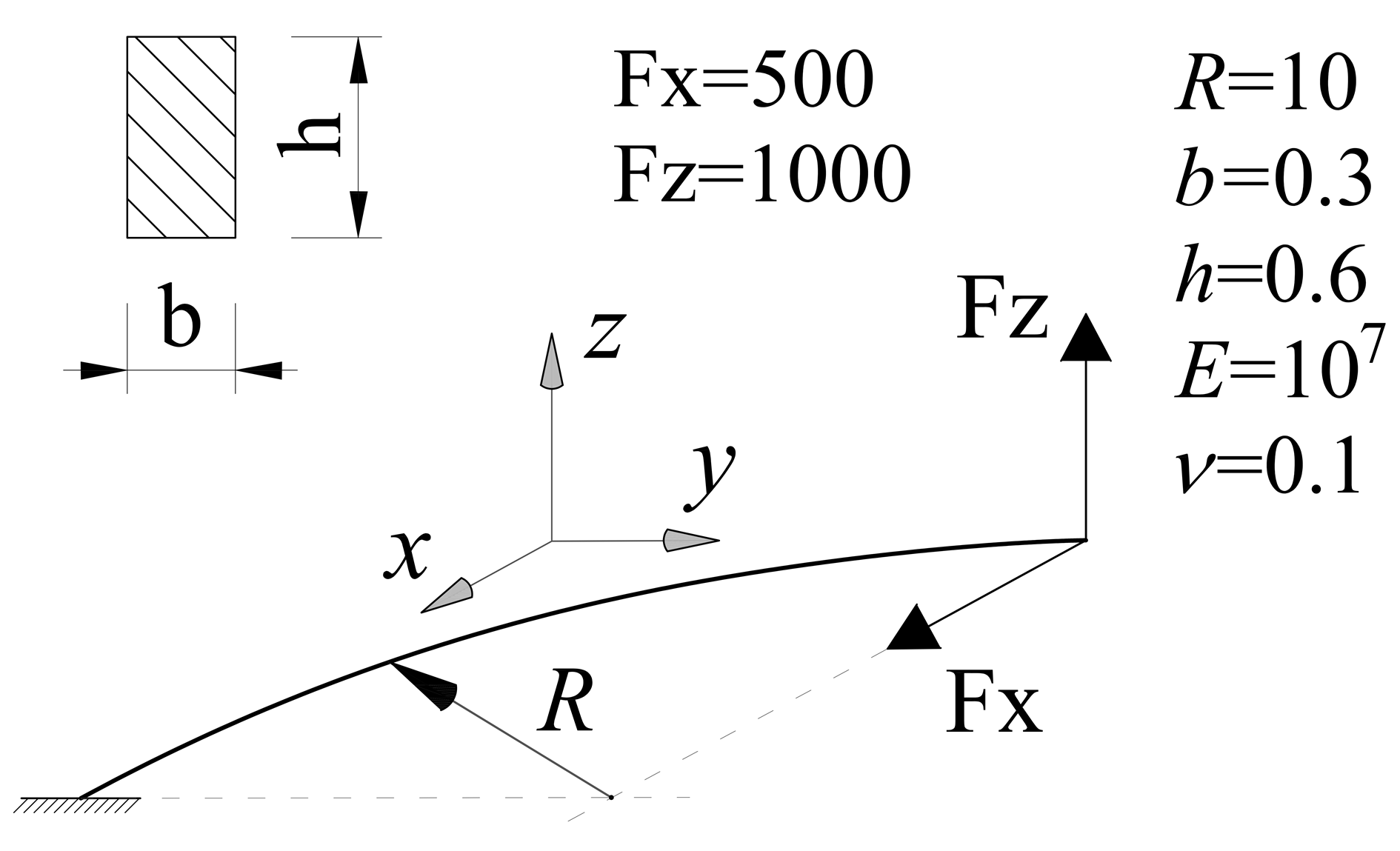}\centering
	\caption{Path-independence. Geometry and load. }
	\label{fig:path disposition}
\end{figure}

Path-independence of the solution can be analyzed in various ways. Some authors simply apply different sizes of load increments, \cite{1999jelenica}, while the others change the order of the applied load \cite{2014meier}. Here, we employ the latter approach. A quarter-circle cantilever beam is loaded with two forces at the free end, as shown in Fig.~\ref{fig:path disposition}.   
Three cases of the application of load are considered. First, both forces are applied simultaneously - $SIM$ case. For the other two cases, the forces are applied successively, one for $0<LPF<0.5$ and the other for $0.5<LPF<1$. The case when the $\iv{F}{}{X}$ is applied first is designated with $SUCXZ$ while the other case is marked with $SUCZX$. 

\begin{figure}
	\includegraphics[width=\linewidth]{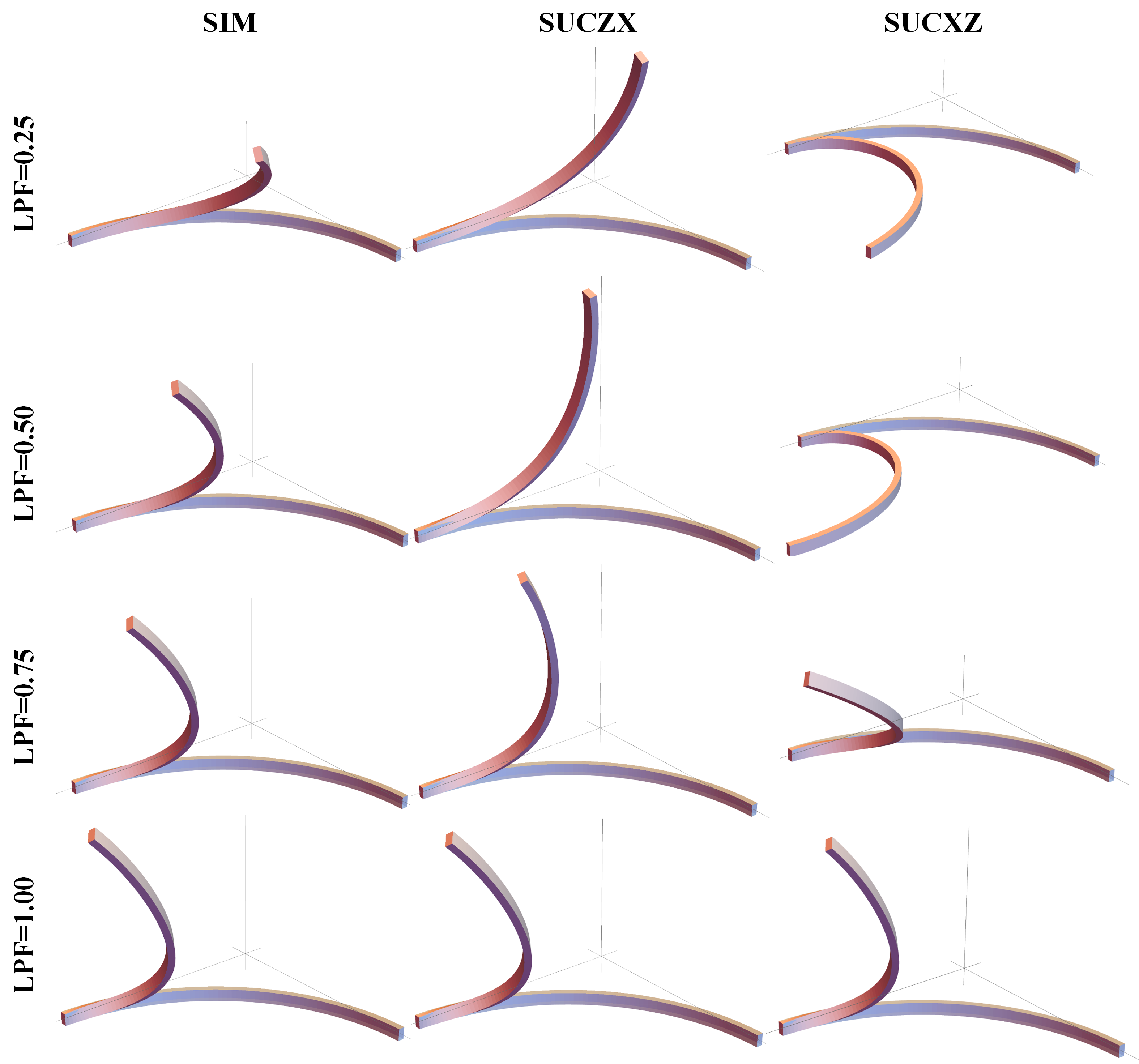}\centering
	\caption{Path-independence. Deformed configurations of the beam for three different loading orders and four values of $LPF$s.}
	\label{fig:path ind config}
\end{figure}

\begin{figure}
	\includegraphics[width=8 cm]{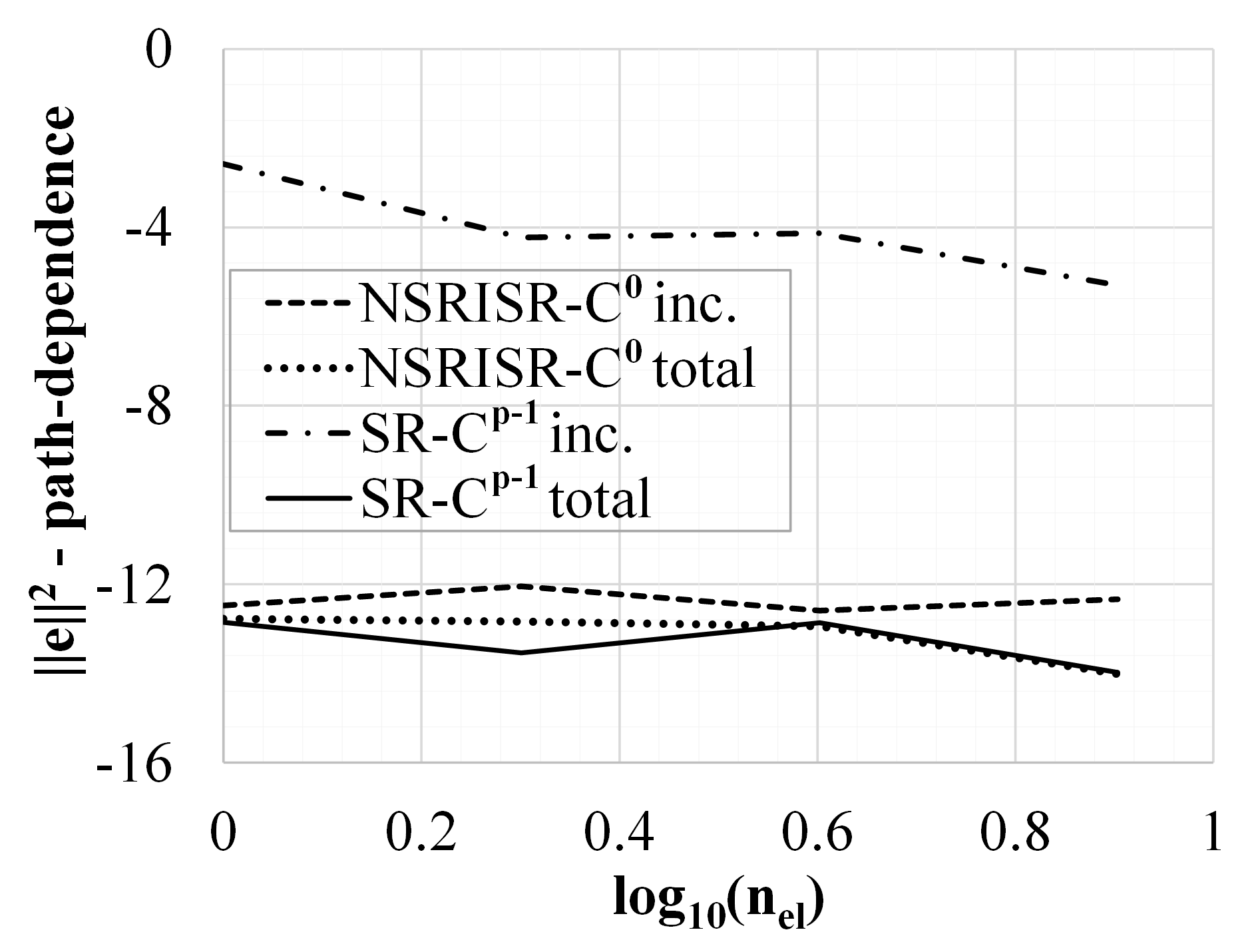}\centering
	\caption{Path-independence. Relative $L^2$-difference of the final configurations calculated by the $SUCZX$ and the $SIM$ load orders with respect to the number of elements. The results are displayed for the $NSRISR-C^{0}$ and $SR-C^{p-1}$ formulations using the incremental and total algorithms for the basis update. }
	\label{fig:path ind comp}
\end{figure}

The deformed beam configurations for all three load order cases and the four characteristic $LPF$s are shown in Fig.~\ref{fig:path ind config}.
Apparently, all load cases yield similar final configurations in visual terms, but each with a different deformation history. Next, the relative $L^2$-error norms for the SIM and SUCZX loading orders are observed for $LPF=1$ using \eqqref{eq: l2 definition}. Two formulations are considered, $SR-C^{p-1}$ and $NSRISR-C^{0}$, and two algorithms for the basis update, incremental and total. The results for quartic NURBS are shown in Fig.~\ref{fig:path ind comp}.
Our simulations show that of these four combinations, only the $SR-C^{p-1}$ formulation with the incremental update of rotations is path-dependent. As expected, when the unstressed configuration is used as the reference one for the update of the basis, the solution is path-independent \cite{1999crisfielda}. An important fact is the confirmation that the $NSRISR-C^{0}$ formulation with incremental update of rotations is path-independent \cite{2014meier}. For all the three considered formulations that are path-independent, the relative $L^2$-error norms are zero up to the machine precision.

\begin{figure}
	\includegraphics[width=\linewidth]{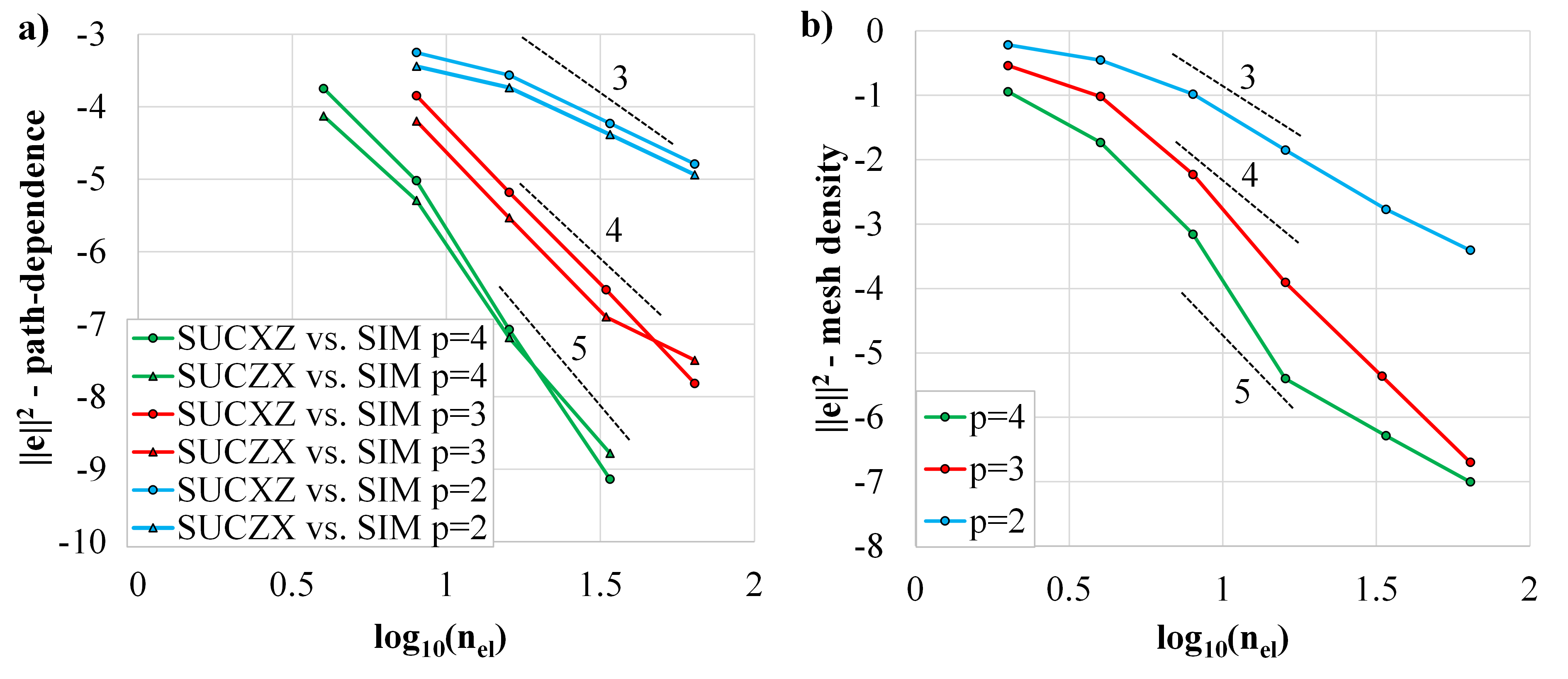}\centering
	\caption{Path-independence. a) Relative $L^2$-error norm for $SR-C^{p-1}$ formulation due to the path-dependence. b) Convergence with respect to the mesh density. }
	\label{fig:path ind converg}
\end{figure}

Regarding the $SR-C^{p-1}$ formulation with incremental update of rotations, the relative $L^2$-error norm is plotted in Fig.~\ref{fig:path ind converg}a for three NURBS orders. 
The error due to the path-dependence mitigates with the mesh refinement. In fact, the order by which the path-dependent error reduces should be similar to the order of convergence of the discretization error \cite{2014meier}. To test this, 200 quartic elements and the $NSRISR-C^{0}$ formulation are used to obtain a reference solution. The convergence of error for three different NURBS orders are displayed in Fig.~\ref{fig:path ind converg}b. For this graph, results from the $NSRISR-C^{0}$ formulation are used with SIM load order. Nevertheless, the discretization errors are practically the same for all the formulations and load orders. The expected orders of convergence, close to $p+1$, are observed and they are similar to those in Fig.~\ref{fig:path ind converg}a. Furthermore, the comparison of the magnitudes of error in Fig.~\ref{fig:path ind converg}a and Fig.~\ref{fig:path ind converg}b confirms the fact that the path-dependent error can be considered negligible with respect to the discretization error \cite{2014meier}.

In the following, if not explicitly stated otherwise, the $NSRISR-C^{0}$ formulation with incremental update of rotations is exclusively used.

\subsection{Circular ring subjected to twisting}

\begin{figure}
	\includegraphics[width=12cm]{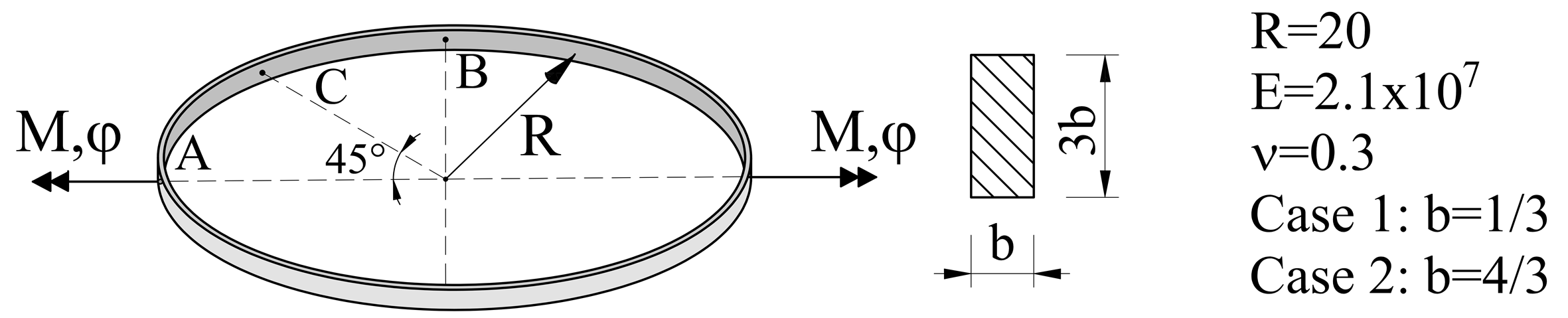}\centering
	\caption{Circular ring subjected to twisting. Load and geometry. }
	\label{fig:Ring1}
\end{figure}

\begin{figure}
	\includegraphics[width=\linewidth]{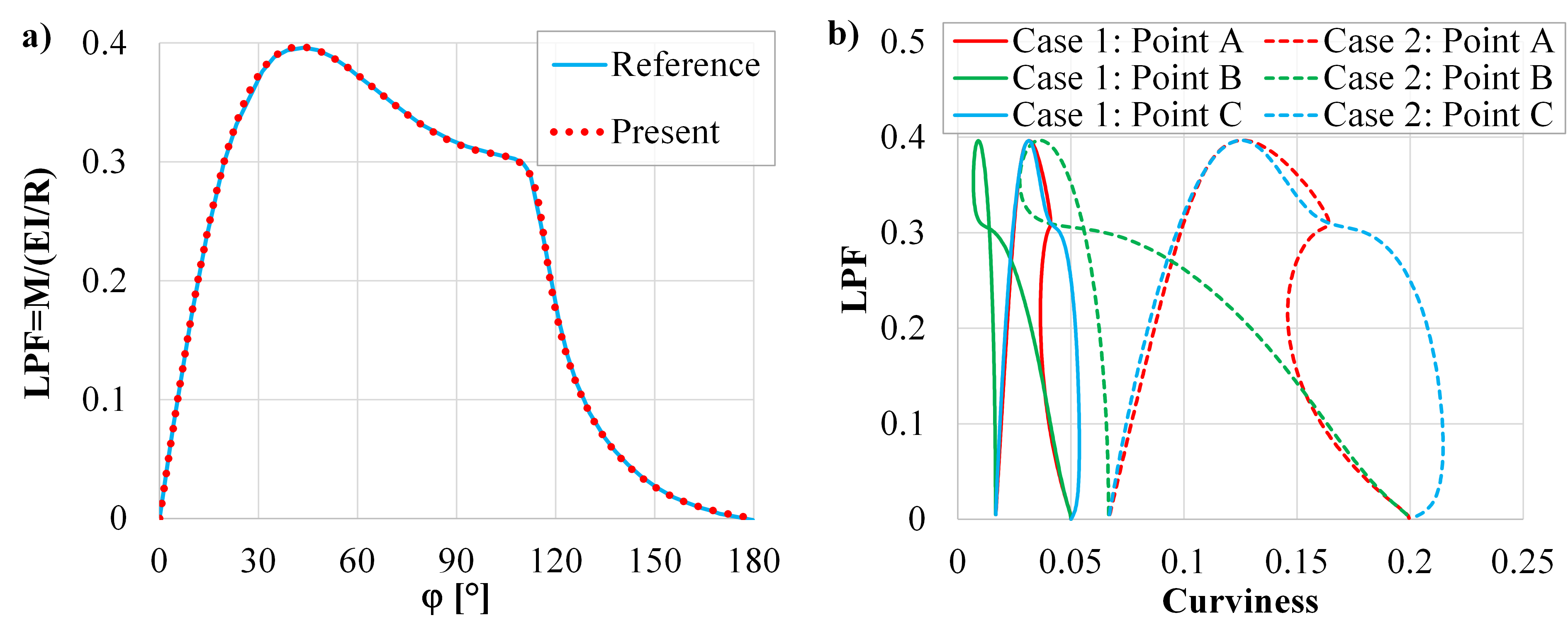}\centering
	\caption{Circular ring subjected to twisting. a) Comparison of LPF vs. external angle of twist. b) Curviness at three points. }
	\label{fig:Ring2}
\end{figure}

This is a well-known example that is frequently considered for the validation of formulations involving finite rotations \cite{2014meier, 2020magisano}. A circular ring is subjected to the symmetrical twisting as shown in Fig.~\ref{fig:Ring1}.
Here, the twist is applied as a pair of external concentrated moments $M=EI/R$. Due to the symmetry of the load and the geometry, only a quarter of the ring is modeled \cite{1992yoshiaki}. In order to obtain converged strains, a dense mesh of 32 quartic elements is used. Furthermore, two different dimensions of the cross section are considered and designated as Case 1 and Case 2, see Fig.~\ref{fig:Ring1}. Case 1 corresponds to the examples that are frequently found in the literature. The dependence of $LPF$ and the external angle of twist on one side of the ring is usually observed for the verification. We have adopted the result from \cite{1996pai} as the reference solution. The obtained result is compared with the reference result given in Fig.~\ref{fig:Ring2}a and excellent agreement can be observed. 
Furthermore, the curviness at three characteristic points is displayed in Fig.~\ref{fig:Ring2}b. For Case 1, the maximum curviness is less than 0.07 and all constitutive models return the same results. 

\begin{figure}
	\includegraphics[width=\linewidth]{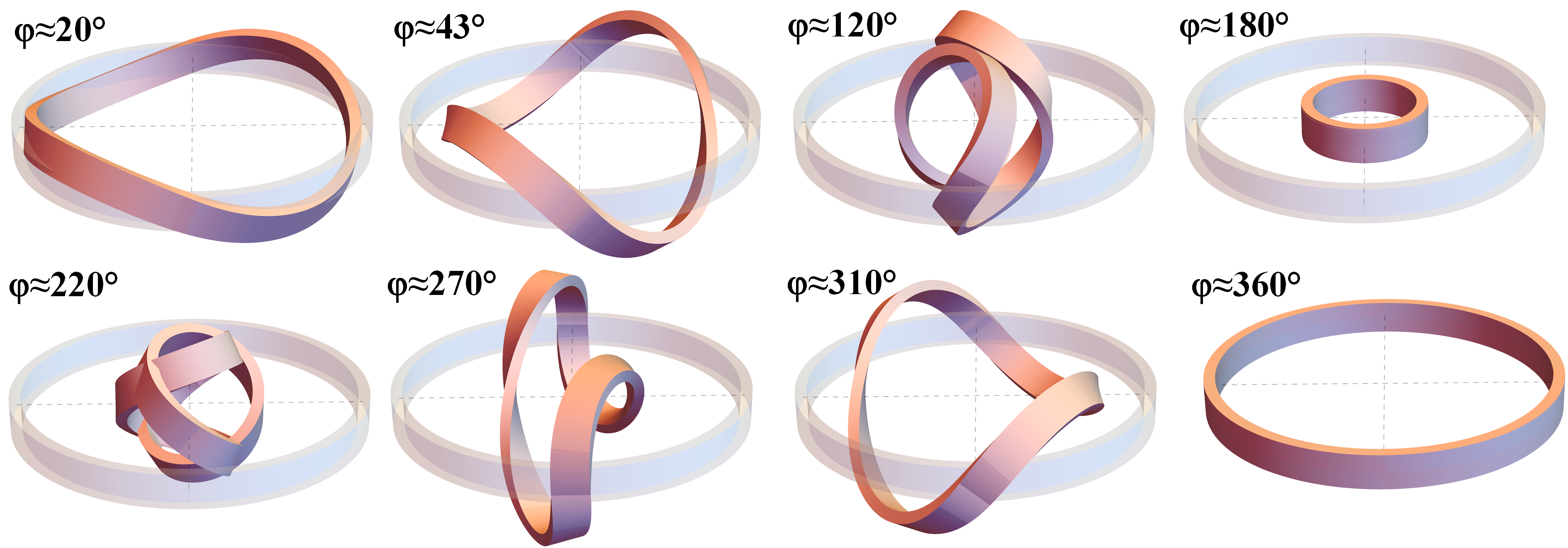}\centering
	\caption{Circular ring subjected to twisting.  Deformed configurations for Case 2. }
	\label{fig:Ring3}
\end{figure}

In order to examine the influence of different constitutive models on a beam with large curviness, Case 2 is now considered in detail. Although the initial curviness for this beam is 1/15, it increases to almost 0.22 during the deformation, Fig.~\ref{fig:Ring2}b, and the beam becomes strongly curved. Furthermore, note that the external twisting of $\varphi=180^{\circ}$ deforms the ring into a smaller ring, with a diameter reduced by a factor of three. Additional application of the external twisting returns the ring into its original configuration for $\varphi=360^{\circ}$, Fig.~\ref{fig:Ring3}.

\begin{figure}
	\includegraphics[width=\linewidth]{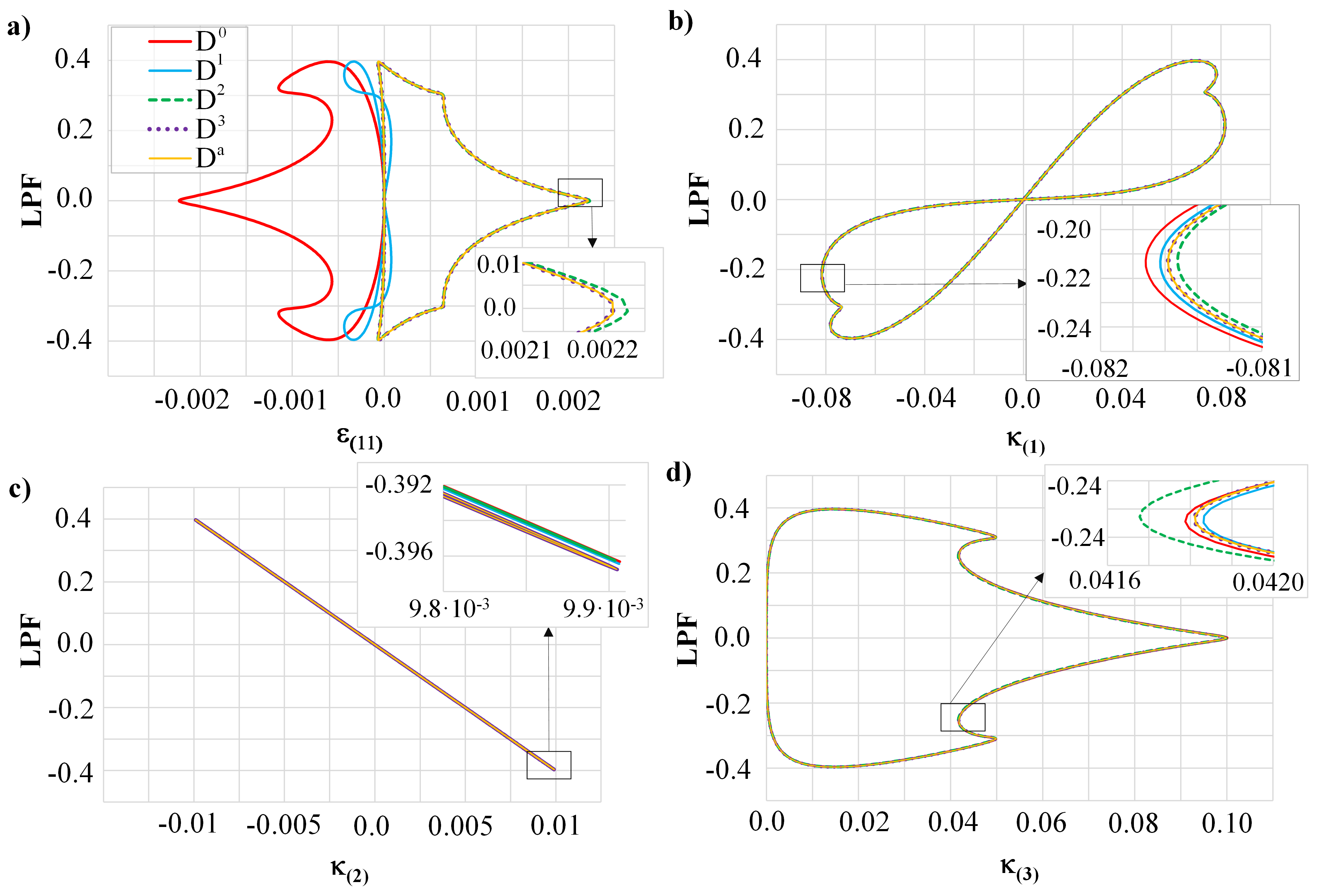}\centering
	\caption{Circular ring subjected to twisting. Reference strains at point A for Case 2. }
	\label{fig:Ring4}
\end{figure}

\begin{figure}
	\includegraphics[width=\linewidth]{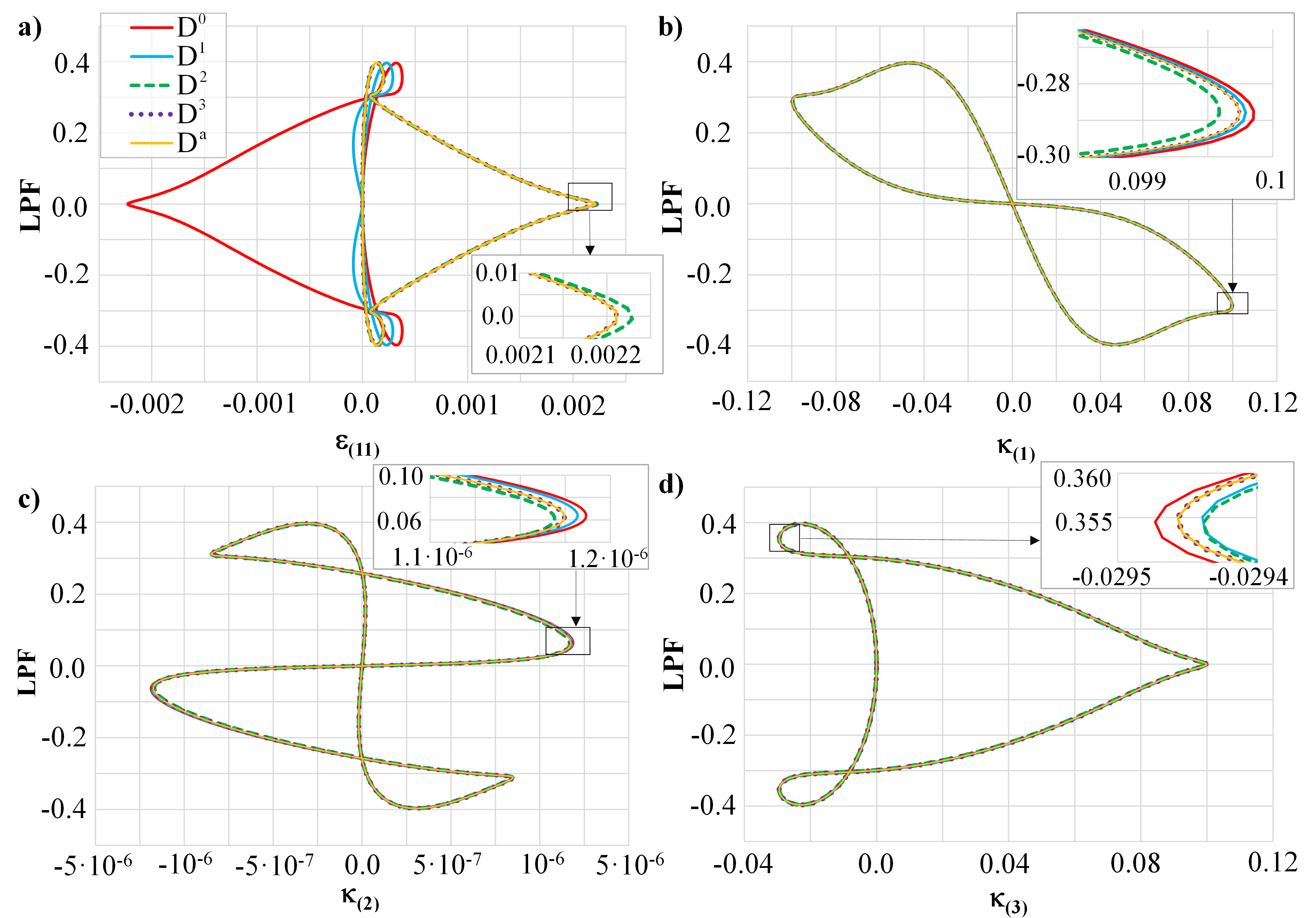}\centering
	\caption{Circular ring subjected to twisting.  Reference strains at point B for Case 2. }
	\label{fig:Ring5}
\end{figure}

In the following, the complete cycle of external twisting is considered ($\varphi=360^{\circ}$) and the reference strains of the beam axis are observed at points A and B, Figs.~\ref{fig:Ring4} and \ref{fig:Ring5}.
The $D^0$ and $D^1$ models return erroneous results for the axial strain. For the other three strain components, the results obtained by the different constitutive models are similar, mostly due to the fact that the maximum curviness of this beam is local. However, a close inspection of the equilibrium paths reveals that the differences exist. This detailed insight suggests that the $D^3$ and $D^a$ models return practically indistinguishable results.

\begin{figure}
	\includegraphics[width=\linewidth]{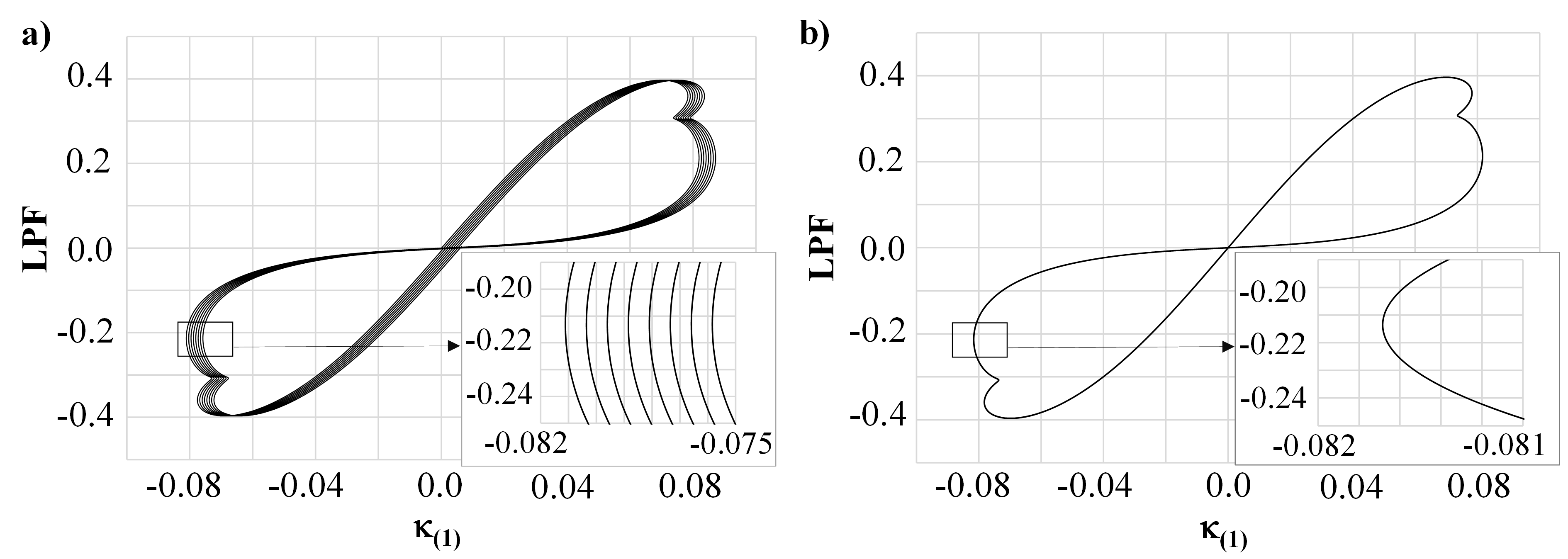}\centering
	\caption{Circular ring subjected to twisting.  Comparison of torsional curvature at point A after eight cycles of twisting for two different formulations: a) $SR-C^0$; b) $NSRISR-C^0$. }
	\label{fig:Ring6}
\end{figure}

The example is suitable for the testing of the path-independence due to the cyclic response of this ring  \cite{2020magisano}. For this purpose, the torsional strain at point A is observed, while the ring is twisted eight times ($\varphi=16\pi$). The results calculated with the $SR-C^0$ and $NSRISR-C^0$ formulations using the incremental update of the basis are given in Fig.~\ref{fig:Ring6}.
This analysis confirms the conclusions from the previous example. The $NSRISR-C^0$ formulation is path-independent while the $SR-C^0$ formulation with incremental update of the basis is not.

\subsection{Straight beam bent to helix}

\begin{figure}
	\includegraphics[width=\linewidth]{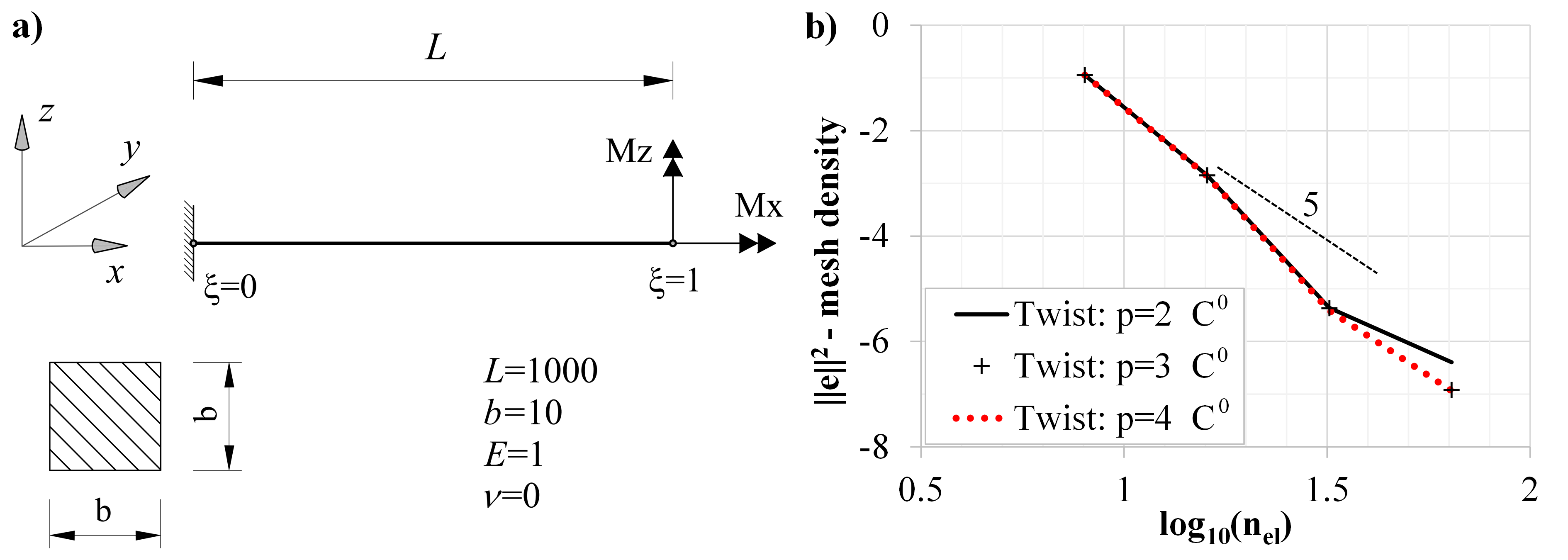}\centering
	\caption{Straight beam bent to helix. a) Geometry and load. b) Convergence of solution with respect to the polynomial degree used for the discretization of the angle of twist.  }
	\label{fig:Helix1}
\end{figure}

As a final example, let us consider the response of an initially straight cantilever loaded with two moments as indicated in Fig.~\ref{fig:Helix1}a. 
The influence of the polynomial used for the approximation of the twist variable is examined first. The beam is loaded with $M_x=M_z=10$, while the displacement of the beam axis is discretized with quartic elements. The relative $L^2$-error norm of the position of the beam axis at the final configuration is calculated by \eqqref{eq: l2 definition}. The reference solution is obtained from an analytical expression for beams with small curvature, proposed in \cite{2015meiera}. Convergence of the $D^1$ model towards the reference solution is shown in Fig.~\ref{fig:Helix1}b. We can observe that the polynomial order used for the discretization of the angle of twist does not have significant influence. In fact, all the considered approximative functions return similar results with the exception of quadratic polynomial for the densest mesh. Moreover, the influence of the polynomial order of the twist variable is negligible for all present numerical results.

\begin{figure}
	\includegraphics[width=\linewidth]{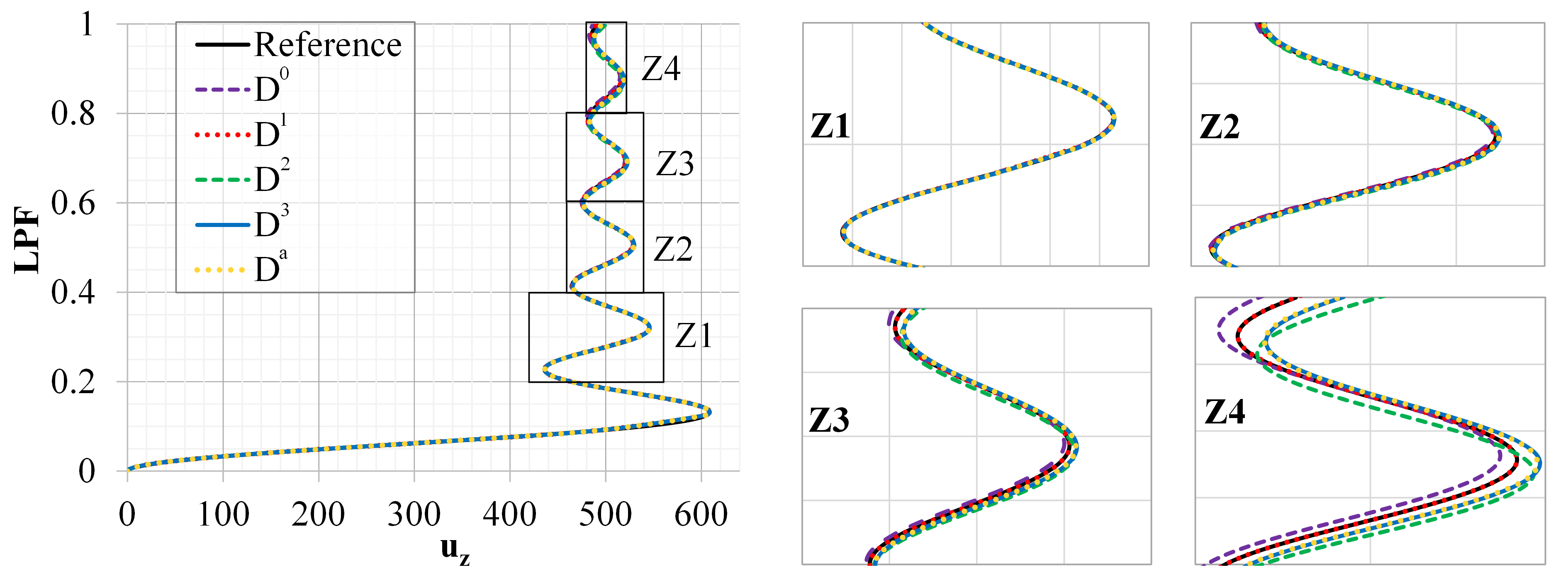}\centering
	\caption{Straight beam bent to helix. Comparison of the tip displacement along the $z$-axis for different constitutive models. Zoomed parts of the equilibrium paths are shown on the right.}
	\label{fig:Helix2}
\end{figure}

Next, the beam is discretized with 40 quartic elements and loaded with $M_x=M_z=20$. The tip displacement along the $z$-axis is plotted in Fig.~\ref{fig:Helix2} for different constitutive models. 
In order to make close inspection, parts of the equilibrium path for $LPF>0.2$ are enlarged. As the load increases, the differences between models become visible. This is due to the fact that the maximum curviness for this example is $Kh\approx0.34$. The $D^1$ model fully corresponds to the reference analytical solution of the small-curvature beam model. The results obtained by the $D^3$ and $D^a$ model are practically identical. The error of the $D^0$ and $D^2$ models increases with the $LPF$.

\begin{figure}
	\includegraphics[width=\linewidth]{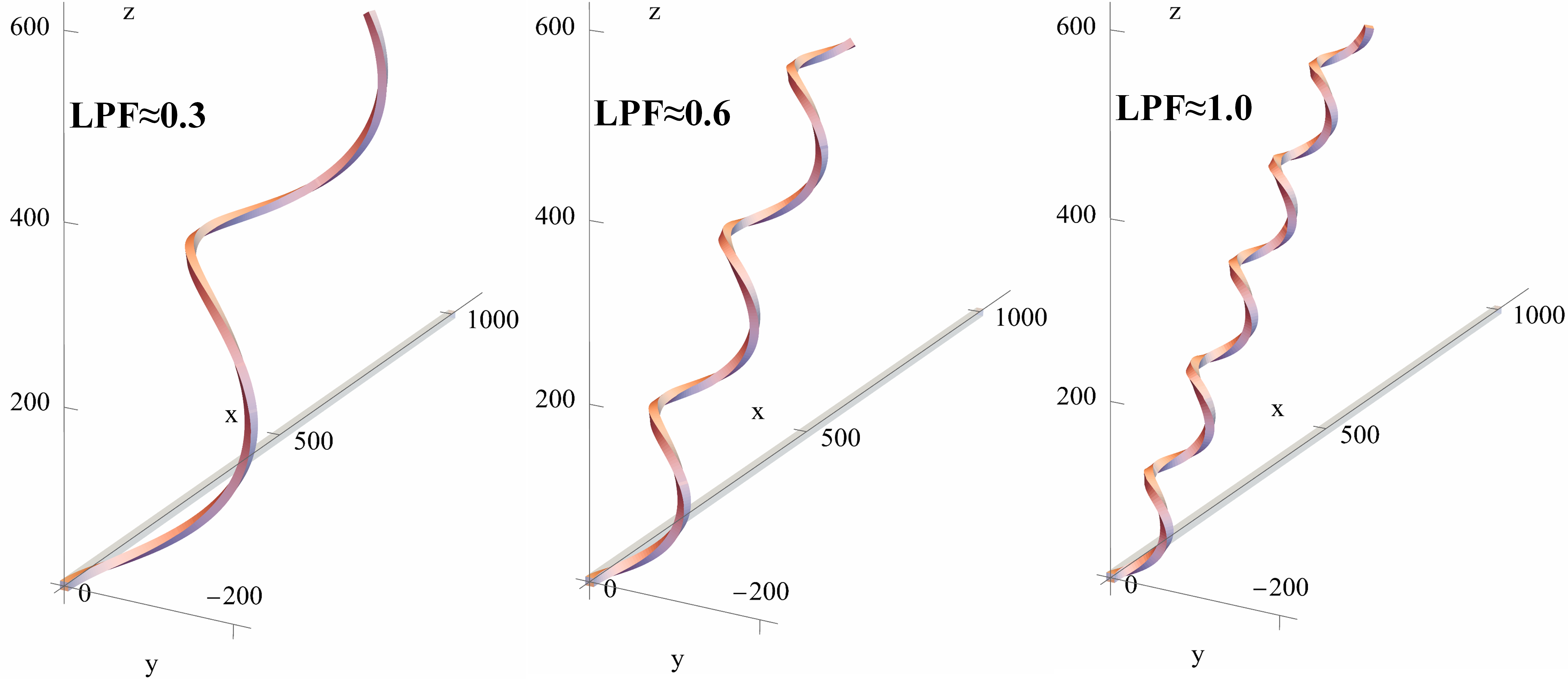}\centering
	\caption{Straight beam bent to helix. Deformed configurations for different values of $LPF$. }
	\label{fig:Helix3}
\end{figure}

\begin{figure}
	\includegraphics[width=\linewidth]{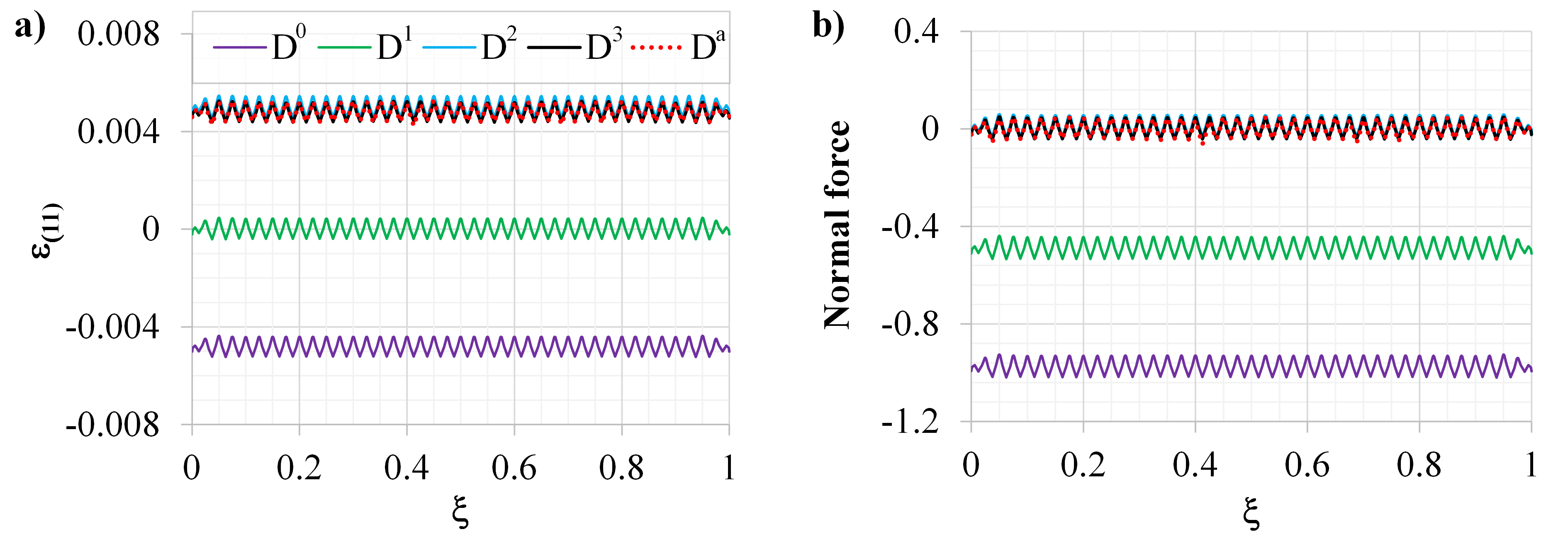}\centering
	\caption{Straight beam bent to helix. a) Distributions of the axial strain of beam axis for different constitutive models. b) Distributions of the normal force for different constitutive models. }
	\label{fig:Helix4}
\end{figure}

The deformed configurations of the beam in Fig.~\ref{fig:Helix3} reveal the complexity of its response.
The beam deforms into a helix, and the equilibrium requires that the normal force along the whole beam is zero. If the small-curvature beam model is utilized, the axial strain of the beam axis must also be zero. However, an accurate model, such as the one presented here, results in the dilatation of the beam axis, in a manner similar to that in \cite{2021borkovic}. The distribution of the axial strain and the normal force at the final configuration are given in Fig.~\ref{fig:Helix4} for different constitutive models.
The $D^0$ model returns an erroneous sign of the dilatation of the beam axis and non-zero normal force. On the other hand, the $D^1$ model results in zero dilatation and non-zero normal force. It should be noted that this model would result in zero normal force if the reduced constitutive matrix is used for the post-processing of the section forces \cite{2021borkovic}. Here, the full constitutive relation is utilized for the calculation of section forces \cite{2018radenkovicb}. Finally, the $D^2$, $D^3$, and $D^a$ models return similar results, with extensional axial strain and near-zero normal force. Again, the $D^3$ and $D^a$ models are fully aligned while the $D^2$ model differs slightly.

\begin{figure}
	\includegraphics[width=\linewidth]{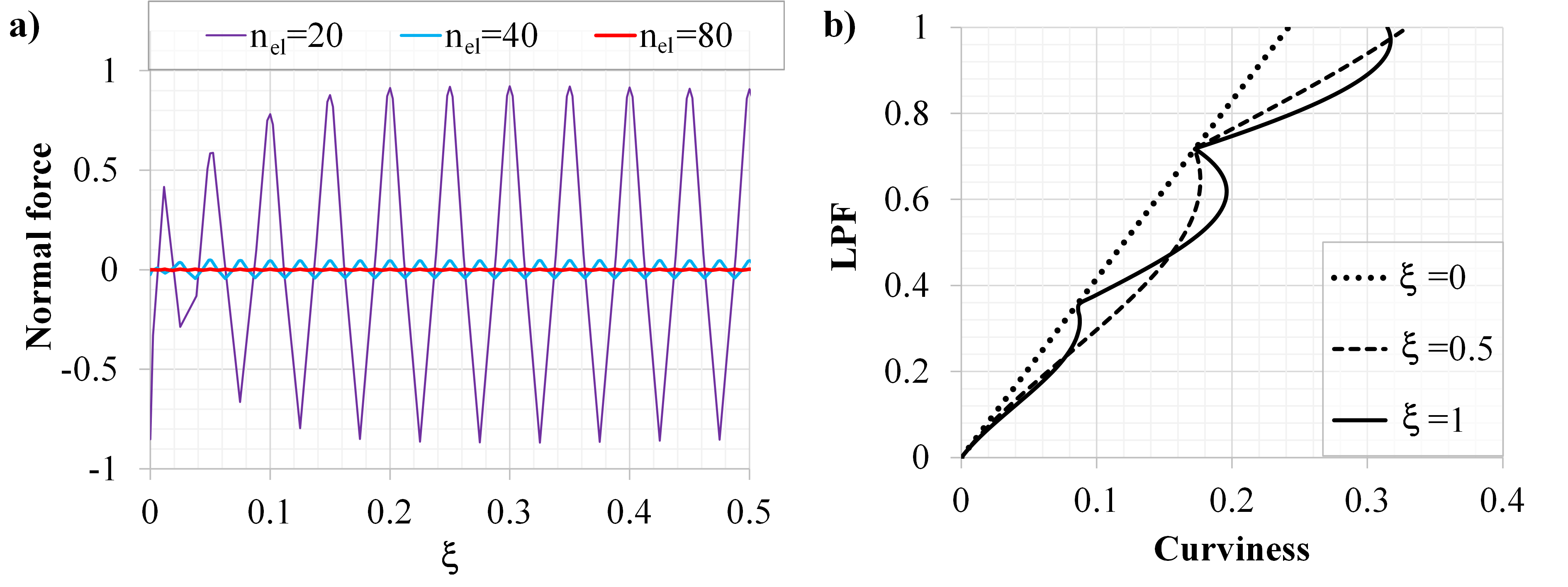}\centering
	\caption{Straight beam bent to helix.  a) Distributions of normal force for different numbers of elements. b) Curviness at the start $(\xi=0)$, at the middle $(\xi=0.5)$, and at the end $(\xi=1)$ of the beam vs. LPF.}
	\label{fig:Helix5}
\end{figure}

Evidently, both the axial strain and the normal force oscillate strongly around the exact value. To examine this effect, the normal forces for three different mesh densities are given in Fig.~\ref{fig:Helix5}a.
The oscillation of these quantities reduces with the mesh density which can indicate the presence of membrane locking. 

The development of the curviness at three characteristics points is shown in Fig.~\ref{fig:Helix5}b. It is interesting to note that the curviness at the clamped end increases monotonically while it is not the case at the other positions. This is due to the fact that the curviness at the clamped cross section varies solely due to the change in curvature. For the other points, the cross section rotates and the curviness exhibits a more complex behavior.

\section{Conclusions}

The first truly geometrically exact isogeometric formulation of a spatial Bernoulli-Euler (BE) beam is presented. The rigorous metric of the BE beam is utilized consistently for the derivation of the weak form of equilibrium. The introduction of the full beam metric gave a higher-order accurate BE beam formulation. The exact constitutive relation is employed for the derivation of four simplified models which are compared via numerical examples. Moreover, by the implementation of the Nodal Smallest Rotation Smallest Rotation Interpolation (NSRISR) mapping, an objective and path-independent formulation is obtained. 

It is confirmed that the Smallest Rotation (SR) mapping results in a non-objective formulation, the error of which reduces with the mesh density. Additionally, it is shown that both the NSRISR and the SR algorithm benefit from a $C^0$ interelement continuity of the twist variable. 

In order to obtain the correct geometric stiffness matrix, both the internal and the external virtual power are rigorously varied with respect to the unknown metric. The angle of twist requires special attention, since it has one part that depends on the geometry and its variation must be performed consistently.

The presented results suggest that, in order to correctly determine the strains of a strongly curved beam, a higher-order accurate computational model must be employed. For the beams with small curviness the simple decoupled equations return reasonably accurate results. As the curviness increases, its influence becomes noticeable and a more involved model is required. A simple yet effective solution to improve the accuracy of small-curvature formulations should include the exact nonlinear distribution of strain and stress in the post-processing phase \cite{2021borkovic}.

Future research into the proposed formulation will deal with the dynamics and material non-linear analysis of beams and their interactions with shells.

\section*{Acknowledgments}

During this work, our beloved colleague and friend, Professor Gligor Radenković (1956-2019), passed away. The first author acknowledges that his unprecedented enthusiasm and love for mechanics were crucial for much of his previous, present, and future research.

We acknowledge the support of the Austrian Science Fund (FWF): M 2806-N.

\section*{Appendix A. Geometric stiffness matrix}
\setcounter{equation}{0}
\renewcommand\theequation{A\arabic{equation}}

Since the variations of the base vectors with respect to the metric are:
\begin{equation}
\label{eq: ap1 variation of base vector g_alpha}
\delta \iv{g}{}{m} = \delta \iv{v}{}{,m} \Delta t,
\end{equation}
the variations of the reference strains with respect to the metric are given by:
\begin{equation}
\label{eq: ap2 variations of reference strains}
\begin{aligned}
\delta \ii{d}{}{11} &= \delta \left( \iv{g}{}{1} \cdot \iv{v}{}{,1} \right) = \delta \iv{v}{}{,1} \cdot \iv{v}{}{,1}, \\
\delta \imd{\kappa}{}{1} &= \delta \left[ \ii{K}{}{2} \left( \iv{g}{}{2} \cdot \iv{v}{}{,1} \right) + \ii{K}{}{3} \left( \iv{g}{}{3} \cdot \iv{v}{}{,1} \right) + \ii{\omega}{}{,1}\right] \\
&= \delta \ii{K}{}{2} \left( \iv{g}{}{2} \cdot \iv{v}{}{,1} \right) +\ii{K}{}{2} \left( \delta \iv{v}{}{,2} \cdot \iv{v}{}{,1} \right) + \delta \ii{K}{}{3} \left( \iv{g}{}{3} \cdot \iv{v}{}{,1} \right) + \ii{K}{}{3} \left( \delta \iv{v}{}{,3} \cdot \iv{v}{}{,1} \right) + \delta \ii{\omega}{}{,1},\\
\delta \imd{\kappa}{}{2} &= \delta \left[ -\iv{g}{}{3} \cdot \left( \iv{v}{}{,11} - \ii{\Gamma}{1}{11} \iv{v}{}{,1}\right) + \ic{K}{}{3} \omega \right] \\
&= -\delta \iv{v}{}{,3} \cdot \left( \iv{v}{}{,11} - \ii{\Gamma}{1}{11} \iv{v}{}{,1}\right) + \delta \ii{\Gamma}{1}{11} \iv{g}{}{3} \cdot  \iv{v}{}{,1}+ \delta \ic{K}{}{3} \omega + \ic{K}{}{3} \delta \omega,\\
\delta \imd{\kappa}{}{3} &= \delta \left[ \iv{g}{}{2} \cdot \left( \iv{v}{}{,11} - \ii{\Gamma}{1}{11} \iv{v}{}{,1}\right) - \ic{K}{}{2} \omega \right] \\
&= \delta \iv{v}{}{,2} \cdot \left( \iv{v}{}{,11} - \ii{\Gamma}{1}{11} \iv{v}{}{,1}\right) - \delta \ii{\Gamma}{1}{11} \iv{g}{}{2} \cdot \iv{v}{}{,1} - \delta \ic{K}{}{2} \omega - \ic{K}{}{2} \delta \omega.
\end{aligned}
\end{equation}
In the following, it is assumed that $\Delta t = 1$, for brevity. 

The variations of the gradients of velocities with respect to the kinematics are easily computed from \eqqref{eq:461}:
\begin{equation}
\label{eq: ap3 variations of reference strains}
\begin{aligned}
\delta \iv{v}{}{,2} &= -\frac{1}{g} \left( \iv{g}{}{2} \cdot \delta \iv{v}{}{,1} \right) \iv{g}{}{1} + \iv{g}{}{3} \delta \omega, \\
\delta \iv{v}{}{,3} &= -\frac{1}{g} \left( \iv{g}{}{3} \cdot \delta \iv{v}{}{,1} \right) \iv{g}{}{1} - \iv{g}{}{2} \delta \omega,
\end{aligned}
\end{equation}
and from \eqqref{eq: gradients v21 v31}:
\begin{equation}
\label{eq: var gradients v21 v31}
\begin{aligned}
\delta \iv{v}{}{,21} &= -\frac{1}{g^*} \bigg\{ \left( \idef{\Gamma}{1}{11} \ivdef{g}{}{1} +\icdef{K}{}{3} \ivdef{g}{}{2} -\icdef{K}{}{2} \ivdef{g}{}{3} \right) \left( \ivdef{g}{}{2} \cdot \delta \iv{v}{}{,1} \right)  + \left[ \left( \idef{K}{}{1} \ivdef{g}{}{3} -\idef{K}{}{3} \ivdef{g}{}{1} \right) \cdot \delta  \iv{v}{}{,1} \right] \ivdef{g}{}{1} \\
& \;\;\;\; + \left( \ivdef{g}{}{2} \cdot \delta \iv{v}{}{,11} \right) \ivdef{g}{}{1} -2 ~ \idef{\Gamma}{1}{11} \left( \ivdef{g}{}{2} \cdot \delta \iv{v}{}{,1} \right) \ivdef{g}{}{1}  \bigg\} + \left( \idef{K}{}{2} \ivdef{g}{}{1} - \idef{K}{}{1} \ivdef{g}{}{2}\right) \delta \omega + \ivdef{g}{}{3} \delta \omega _{,1} ,\\
\delta \iv{v}{}{,31} &= -\frac{1}{g^*} \bigg\{  \left( \idef{\Gamma}{1}{11} \ivdef{g}{}{1} +\icdef{K}{}{3} \ivdef{g}{}{2} -\icdef{K}{}{2} \ivdef{g}{}{3} \right) \left( \ivdef{g}{}{3} \cdot \delta \iv{v}{}{,1} \right)  + \left[ \left( \idef{K}{}{2} \ivdef{g}{}{1} - \idef{K}{}{1} \ivdef{g}{}{2}\right) \cdot \delta \iv{v}{}{,1} \right] \ivdef{g}{}{1} \\
& \;\;\;\; + \left( \ivdef{g}{}{3} \cdot \delta \iv{v}{}{,11} \right) \ivdef{g}{}{1} - 2~ \idef{\Gamma}{1}{11} \left( \ivdef{g}{}{3} \cdot \delta \iv{v}{}{,1} \right)\ivdef{g}{}{1} \bigg\} + \left( \idef{K}{}{3} \ivdef{g}{}{1} - \idef{K}{}{1} \ivdef{g}{}{3}\right) \delta \omega - \ivdef{g}{}{2} \delta \omega _{,1} .
\end{aligned}
\end{equation}
Variations of the curvature components with respect to the metric are:
\begin{equation}
\label{eq: ap4 variations of reference strains}
\begin{aligned}
\delta \ic{K}{}{2} &= \delta \left( - \iv{g}{}{1,1} \cdot \iv{g}{}{3} \right)= -\iv{g}{}{3} \cdot \left( \delta \iv{v}{}{,11} - \ii{\Gamma}{1}{11} \delta \iv{v}{}{,1}\right) + \ic{K}{}{3} \delta \omega, \\
\delta \ic{K}{}{3} &= \delta \left( \iv{g}{}{1,1} \cdot \iv{g}{}{2} \right)= \iv{g}{}{2} \cdot \left( \delta \iv{v}{}{,11} - \ii{\Gamma}{1}{11} \delta \iv{v}{}{,1}\right) - \ic{K}{}{2} \delta \omega, \\
\delta \ii{K}{}{2} &= \delta \left(\frac{\ic{K}{}{2}}{g} \right)= \frac{1}{g} \left( \delta \ic{K}{}{2} - \frac{2}{g} \ic{K}{}{2} ~ \iv{g}{}{1} \cdot \delta \iv{v}{}{,1} \right), \\
\delta \ii{K}{}{3} &= \delta \left(\frac{\ic{K}{}{3}}{g} \right)= \frac{1}{g} \left( \delta \ic{K}{}{3} - \frac{2}{g} \ic{K}{}{3} ~ \iv{g}{}{1} \cdot \delta \iv{v}{}{,1} \right),
\end{aligned}
\end{equation}
while the variation of the Christoffel symbol is:
\begin{equation}
\label{eq: ap6 variations of reference strains}
\begin{aligned}
\delta \ii{\Gamma}{1}{11} &= \delta \left(\frac{\iv{g}{}{1,1} \cdot \iv{g}{}{1}}{g} \right)= \frac{1}{g} \left[ \iv{g}{}{1} \cdot \left( \delta \iv{v}{}{,11} - \ii{\Gamma}{1}{11} \delta \iv{v}{}{,1}\right) + \left( \ic{K}{}{3} \iv{g}{}{2} - \ic{K}{}{2} \iv{g}{}{3} \right) \cdot \delta \iv{v}{}{,1} \right].
\end{aligned}
\end{equation}
Although the twist angle is adopted as kinematic quantity, it has one part that is pure geometry. Therefore, we must also vary it with respect to the metric:
\begin{equation}
\label{eq: ap7 variations of reference strains}
\begin{aligned}
\delta \omega &= \delta \left(\iv{g}{}{3} \cdot \iv{v}{}{,2} \right) = \frac{1}{g} \left( \iv{g}{}{3} \cdot \delta \iv{v}{}{,1} \right) \left( \iv{g}{}{2} \cdot  \iv{v}{}{,1} \right) = \frac{1}{g} ~ \iv{v}{}{,1} \cdot \left(\iv{g}{}{2} \otimes \iv{g}{}{3} \right) \delta \iv{v}{}{,1}.
\end{aligned}
\end{equation}
\begin{remark}
It is interesting to note that, since the angular velocity can be written as $\omega = \iv{g}{}{3} \cdot \iv{v}{}{,2}$ or as $\omega = - \iv{g}{}{2} \cdot \iv{v}{}{,3}$, its variation with respect to the metric can be written in two ways. The first one is given with \eqqref{eq: ap7 variations of reference strains} while the other one is:
\begin{equation}
\label{eq: ap7 variations of reference strains2}
\begin{aligned}
\delta \omega = \delta \left(- \iv{g}{}{2} \cdot \iv{v}{}{,3} \right) = - \frac{1}{g} \left( \iv{g}{}{2} \cdot \delta \iv{v}{}{,1} \right) \left( \iv{g}{}{3} \cdot  \iv{v}{}{,1} \right) = - \frac{1}{g} ~ \iv{v}{}{,1} \cdot \left(\iv{g}{}{3} \otimes \iv{g}{}{2} \right) \delta \iv{v}{}{,1}.
\end{aligned}
\end{equation} 
Both representation are valid and can be used for the derivation of the geometric stiffness. However, the choice of the representation for $\delta \omega$ must be consistently applied. Its inconsistent use is actually the source of error in the reference \cite{2017radenkovic}. Furthermore, although the both representations of the angular velocity return the same value, their variations in Eqs.~\eqref{eq: ap7 variations of reference strains} and \eqref{eq: ap7 variations of reference strains2} are not the same, since the variation is performed with respect to the different parameters.
\end{remark}

These expressions are sufficient for the variations of the bending curvatures. For the torsional term, the variation of the gradient of the angular velocity with respect to the metric is required:
\begin{equation}
\label{eq: ap8 variations of reference strains}
\begin{aligned}
\delta \ii{\omega}{}{,1} &= \delta \left(\iv{g}{}{3} \cdot \iv{v}{}{,2} \right) _{,1} = \delta \left( \iv{g}{}{3,1} \cdot \iv{v}{}{,2} + \iv{g}{}{3} \cdot \iv{v}{}{,21} \right) = \delta \iv{v}{}{,31} \cdot \iv{v}{}{,2} + \delta \iv{v}{}{,3} \cdot \iv{v}{}{,21}.
\end{aligned}
\end{equation}
By inserting Eqs.~\eqref{eq:461}, \eqref{eq: gradients v21 v31}, \eqref{eq: ap3 variations of reference strains}, and \eqref{eq: var gradients v21 v31} into \eqqref{eq: ap8 variations of reference strains}, we obtain:
\begin{equation}
\label{eq: ap10 variations of reference strains}
\begin{aligned}
\delta \omega _{,1} &= \frac{1}{g} ~ \iv{v}{}{,1} \left[ -2 ~ \ii{\Gamma}{1}{11} \left( \iv{g}{}{2} \otimes \iv{g}{}{3} \right)  + \ii{K}{}{1} \left( \iv{g}{}{3} \otimes \iv{g}{}{3} \right) - \ii{K}{}{1} \left( \iv{g}{}{2} \otimes \iv{g}{}{2} \right) +  \ii{K}{}{2} \left( \iv{g}{}{2} \otimes \iv{g}{}{1} \right) \right.\\
&\left. \;\;\;\; - \ii{K}{}{3} \left( \iv{g}{}{1} \otimes \iv{g}{}{3} \right) \right] \delta \iv{v}{}{,1} + \frac{1}{g} ~ \iv{v}{}{,1} \cdot \left( \iv{g}{}{2} \otimes \iv{g}{}{3} \right) \delta \iv{v}{}{,11} + \frac{1}{g} ~ \iv{v}{}{,11} \cdot \left( \iv{g}{}{2} \otimes \iv{g}{}{3} \right) \delta \iv{v}{}{,1}.
\end{aligned}
\end{equation}
Finally, by the insertion of Eqs.~\eqref{eq: ap3 variations of reference strains}, \eqref{eq: ap4 variations of reference strains}, \eqref{eq: ap6 variations of reference strains}, \eqref{eq: ap7 variations of reference strains}, and \eqref{eq: ap10 variations of reference strains} into \eqqref{eq: ap2 variations of reference strains}, the variations of curvature strain rates with respect to the geometry are:
\begin{equation}
\label{eq: ap11 variations of reference strains}
\begin{aligned}
\delta \imd{\kappa}{}{1} &=  \frac{1}{g} ~ \iv{v}{}{,1} \cdot \left[ - \ii{\Gamma}{1}{11} \left( \iv{g}{}{2} \otimes \iv{g}{}{3} + \iv{g}{}{3} \otimes \iv{g}{}{2} \right) + \ii{K}{}{1} \left( \iv{g}{}{3} \otimes \iv{g}{}{3} - \iv{g}{}{2} \otimes \iv{g}{}{2}  \right)    \right.\\
&\left. \;\;\;\; - \ii{K}{}{2} \left( \iv{g}{}{2} \otimes \iv{g}{}{1} +  \iv{g}{}{1} \otimes \iv{g}{}{2}  \right) - 2 \ii{K}{}{3} \left( \iv{g}{}{3} \otimes \iv{g}{}{1} + \iv{g}{}{1} \otimes \iv{g}{}{3} \right)  \right] \delta \iv{v}{}{,1} \\
& \;\;\;\; + \frac{1}{g} ~ \iv{v}{}{,1} \cdot \left( \iv{g}{}{3} \otimes \iv{g}{}{2} \right) \delta \iv{v}{}{,11} + \frac{1}{g} ~ \iv{v}{}{,11} \cdot \left( \iv{g}{}{2} \otimes \iv{g}{}{3} \right) \delta \iv{v}{}{,1}, \\
\delta \imd{\kappa}{}{2} &=  \frac{1}{g} ~ \iv{v}{}{,1} \cdot \left[ - \ii{\Gamma}{1}{11} \left( \iv{g}{}{1} \otimes \iv{g}{}{3} + \iv{g}{}{3} \otimes \iv{g}{}{1} \right) - \ic{K}{}{2} \left( \iv{g}{}{3} \otimes \iv{g}{}{3} \right)    \right.\\
&\left. \;\;\;\; + \ic{K}{}{3} \left( \iv{g}{}{2} \otimes \iv{g}{}{3} +  \iv{g}{}{3} \otimes \iv{g}{}{2}  \right) \right] \delta \iv{v}{}{,1} + \frac{1}{g} ~ \iv{v}{}{,1} \cdot \left( \iv{g}{}{3} \otimes \iv{g}{}{1} \right) \delta \iv{v}{}{,11}\\
& \;\;\;\;  + \frac{1}{g} ~ \iv{v}{}{,11} \cdot \left( \iv{g}{}{1} \otimes \iv{g}{}{3} \right) \delta \iv{v}{}{,1} - \iv{v}{}{,1} \cdot \left( \ii{\Gamma}{1}{11} \iv{g}{}{2}\right) \delta \omega - \omega \left( \ii{\Gamma}{1}{11} \iv{g}{}{2} \right)\cdot \delta \iv{v}{}{,1} \\
&\;\;\;\; + \iv{v}{}{,11} \cdot \iv{g}{}{2} \delta \omega + \omega \iv{g}{}{2} \cdot \delta \iv{v}{}{,11} -\omega \ic{K}{}{2} \delta \omega, \\
\delta \imd{\kappa}{}{3} &=  \frac{1}{g} ~ \iv{v}{}{,1} \cdot \left[ \ii{\Gamma}{1}{11} \left( \iv{g}{}{1} \otimes \iv{g}{}{2} + \iv{g}{}{2} \otimes \iv{g}{}{1} \right)     - \ic{K}{}{3} \left( \iv{g}{}{2} \otimes \iv{g}{}{2}  \right) \right] \delta \iv{v}{}{,1} \\
&\;\;\;\;  - \frac{1}{g} ~ \iv{v}{}{,1} \cdot \left( \iv{g}{}{2} \otimes \iv{g}{}{1} \right) \delta \iv{v}{}{,11} - \frac{1}{g} ~ \iv{v}{}{,11} \cdot \left( \iv{g}{}{1} \otimes \iv{g}{}{2} \right) \delta \iv{v}{}{,1}\\
& \;\;\;\;   - \iv{v}{}{,1} \cdot \left( \ii{\Gamma}{1}{11} \iv{g}{}{3} \right) \delta \omega - \omega \left( \ii{\Gamma}{1}{11} \iv{g}{}{3} \right) \cdot \delta \iv{v}{}{,1} + \iv{v}{}{,11} \cdot \iv{g}{}{3} \delta \omega + \omega \iv{g}{}{3} \cdot \delta \iv{v}{}{,11} -\omega \ic{K}{}{3} \delta \omega.
\end{aligned}
\end{equation}
Let us introduce following designations:
\begin{equation}
\label{eq: ap12 variations of reference strains}
\begin{aligned}
\iv{G}{}{11} &= N \textbf{I}_{3\times3}+  \frac{\ii{M}{}{1}}{g} \left[ - \ii{\Gamma}{1}{11} \left( \iv{g}{}{2} \otimes \iv{g}{}{3} + \iv{g}{}{3} \otimes \iv{g}{}{2} \right) + \ii{K}{}{1} \left( \iv{g}{}{3} \otimes \iv{g}{}{3} - \iv{g}{}{2} \otimes \iv{g}{}{2}  \right)    \right.\\
&\left. \;\;\;\; - \ii{K}{}{2} \left( \iv{g}{}{2} \otimes \iv{g}{}{1} +  \iv{g}{}{1} \otimes \iv{g}{}{2}  \right) - 2 \ii{K}{}{3} \left( \iv{g}{}{3} \otimes \iv{g}{}{1} + \iv{g}{}{1} \otimes \iv{g}{}{3} \right)  \right] \\
&\;\;\;\; + \frac{\ii{M}{}{2}}{g}  \left[ - \ii{\Gamma}{1}{11} \left( \iv{g}{}{1} \otimes \iv{g}{}{3} + \iv{g}{}{3} \otimes \iv{g}{}{1} \right) - \ic{K}{}{2} \left( \iv{g}{}{3} \otimes \iv{g}{}{3} \right)    \right.\\
&\left. \;\;\;\; + \ic{K}{}{3} \left( \iv{g}{}{2} \otimes \iv{g}{}{3} +  \iv{g}{}{3} \otimes \iv{g}{}{2}  \right) \right]  \\
&\;\;\;\; + \frac{\ii{M}{}{3}}{g}  \left[ \ii{\Gamma}{1}{11} \left( \iv{g}{}{1} \otimes \iv{g}{}{2} + \iv{g}{}{2} \otimes \iv{g}{}{1} \right) - \ic{K}{}{3} \left( \iv{g}{}{2} \otimes \iv{g}{}{2}  \right) \right], \\
\iv{G}{}{12} &= \frac{\ii{M}{}{1}}{g} \left( \iv{g}{}{3} \otimes \iv{g}{}{2} \right) + \frac{\ii{M}{}{2}}{g} \left( \iv{g}{}{3} \otimes \iv{g}{}{1} \right) - \frac{\ii{M}{}{3}}{g} \left( \iv{g}{}{2} \otimes \iv{g}{}{1} \right), \\
\iv{G}{}{13} &= - \ii{M}{}{2}  \left( \ii{\Gamma}{1}{11} \iv{g}{}{2}\right) - \ii{M}{}{3}  \left( \ii{\Gamma}{1}{11} \iv{g}{}{3} \right), \\
\iv{G}{}{21} &= \frac{\ii{M}{}{1}}{g} \left( \iv{g}{}{2} \otimes \iv{g}{}{3} \right) + \frac{\ii{M}{}{2}}{g} \left( \iv{g}{}{1} \otimes \iv{g}{}{3} \right) - \frac{\ii{M}{}{3}}{g} \left( \iv{g}{}{1} \otimes \iv{g}{}{2} \right), \\
\iv{G}{}{22} &= \textbf{0}_{3\times3}, \\
\iv{G}{}{23} &= \ii{M}{}{2} \iv{g}{}{2} + \ii{M}{}{2} \iv{g}{}{3}, \\
\iv{G}{}{31} &= - \ii{M}{}{2}  \left( \ii{\Gamma}{1}{11} \iv{g}{\text{T}}{2}\right) - \ii{M}{}{3}  \left( \ii{\Gamma}{1}{11} \iv{g}{\text{T}}{3} \right), \\
\iv{G}{}{32} &= \ii{M}{}{2} \iv{g}{\text{T}}{2}+ \ii{M}{}{2} \iv{g}{\text{T}}{3}, \\
\iv{G}{}{33} &= - \ii{M}{}{2} \ic{K}{}{2} - \ii{M}{}{3} \ic{K}{}{3},
\end{aligned}
\end{equation}
which allows us to define the matrix of generalized section forces:
\begin{equation}
\label{eq: apnestoG matrix def}
\ve{G} = 
\begin{bmatrix}
\iv{G}{}{11} & \iv{G}{}{12} & \iv{G}{}{13} \\
\iv{G}{}{21} & \iv{G}{}{22} & \iv{G}{}{23} \\
\iv{G}{}{31} & \iv{G}{}{32} & \iv{G}{}{33}
\end{bmatrix}.
\end{equation}
Let us define the matrix of basis functions $\iv{B}{}{G}$:
\setcounter{MaxMatrixCols}{20}
\begin{equation}
\label{eq: appB}
\begin{aligned}
\iv{B}{}{G} &= 
\begin{bmatrix}
\iv{B}{}{G1} & \iv{B}{}{G2} & ... & \iv{B}{}{GI} & ... & \iv{B}{}{GN} & \iv{B}{\omega}{G1} & \iv{B}{\omega}{G2} & ... & \iv{B}{\omega}{GJ} & ... & \iv{B}{\omega}{GM}
\end{bmatrix}, \\
\iv{B}{}{GI} &=
\begin{bmatrix}
\iv{R}{}{I,1}   \\
\iv{R}{}{I,11}  \\
\textbf{0}_{1\times3}  \\
\end{bmatrix}, \quad
\iv{B}{\omega}{GJ} = 
\begin{bmatrix}
\textbf{0}_{6\times1}  \\
\ii{R}{\omega}{J} 
\end{bmatrix},
\end{aligned}
\end{equation}
which is actually the matrix $\ve{B}$ without the $8^{th}$ row, see \eqqref{eq: vector of reference strains matrix form2}. The difference between these two matrices of the basis functions is due to the fact that the variation of the torsional curvature change with respect to the kinematics depends on the $\ii{\omega}{}{,1}$, cf. Eqs~\eqref{eq: rs1} and \eqref{eq: 1 e=Hw, BL=HB}, while its variation with respect to the metric does not, cf. \eqqref{eq: ap11 variations of reference strains}. The part of the virtual power generated by the known stress and the variation of the strain rate with respect to the metric can now be expressed as:
\begin{equation}
\label{eq: apn1estoG matrix def}
\int_{\xi}^{} \trans{f} \delta \iv{H}{}{} \ve{B} \ivmd{q}{}{} \sqrt{g} \dd{\xi} = \transmd{q} \int_{\xi}^{} \trans{$\iv{B}{}{G}$} \iv{G}{}{} \iv{B}{}{G} \sqrt{g} \dd{\xi} \delta \ivmd{q}{}{} =
\transmd{q} \iv{K}{}{G} \delta \ivmd{q}{}{}.
\end{equation}
Note that the derived geometric stiffness matrix is symmetric. This confirms that the energetically conjugated pairs are correctly adopted. Additionally, the full metric of the BE beam is incorporated in $\iv{K}{}{G}$. 

\section*{Appendix B. Variation of external virtual power with respect to the metric}
\setcounter{equation}{0}
\renewcommand\theequation{B\arabic{equation}}

Let us consider the contribution to the tangent stiffness that comes from the external load. We will focus on the virtual power due to the concentrated moment, since it was the load used in the numerical analysis. This part of the virtual power is:
\begin{equation}
\label{eq: f1}
\delta P_{ext}=-\ve{m} \cdot \delta \bm{\omega}.
\end{equation}
If the angular velocities are varied with respect to the kinematics, we obtain the vector of external load. On the other hand, if we vary the angular velocity with respect to the metric, an addition to the stiffness matrix follows. The variation of angular velocity with respect to the geometry is:
\begin{equation}
\label{eq: f2}
\delta \bm{\omega}  =\delta \left( \ii{\omega}{k}{} \iv{g}{}{k} \right) = \delta \ii{\omega}{k}{} \iv{g}{}{k}  + \ii{\omega}{k}{} \delta \iv{v}{}{,k},
\end{equation}
where the variations of components of angular velocity are:
\begin{equation}
\label{eq: f3}
\begin{aligned}
\delta \ii{\omega}{1}{} &= \delta \left( \frac{1}{\sqrt{g} } \ \omega \right) = -\frac{1}{g^{3/2}} \left( \iv{g}{}{1} \cdot \delta \iv{v}{}{,1} \right) \omega + \frac{1}{g^{3/2}} \iv{v}{}{,1} \left( \iv{g}{}{2} \otimes \iv{g}{}{3} \right) \delta \iv{v}{}{,1}, \\
\delta \ii{\omega}{2}{} &= \delta \left( - \frac{1}{\sqrt{g} } \ \iv{g}{}{3} \cdot \iv{v}{}{,1} \right) = \frac{1}{g^{3/2}} \ \iv{v}{}{,1} \cdot \left( \iv{g}{}{3} \otimes \iv{g}{}{1} + \iv{g}{}{1} \otimes \iv{g}{}{3} \right) \delta \iv{v}{}{,1} + \frac{1}{\sqrt{g}} \ \iv{g}{}{2} \cdot \iv{v}{}{,1} \delta \omega, \\
\delta \ii{\omega}{3}{} &= \delta \left( \frac{1}{\sqrt{g} } \ \iv{g}{}{2} \cdot \iv{v}{}{,1} \right) = - \frac{1}{g^{3/2}} \ \iv{v}{}{,1} \cdot \left( \iv{g}{}{2} \otimes \iv{g}{}{1} + \iv{g}{}{1} \otimes \iv{g}{}{2} \right) \delta \iv{v}{}{,1} + \frac{1}{\sqrt{g}} \ \iv{g}{}{3} \cdot \iv{v}{}{,1} \delta \omega.
\end{aligned}
\end{equation}
By the insertion of Eqs.~\eqref{eq:45}, \eqref{eq: ap3 variations of reference strains}, \eqref{eq: ap7 variations of reference strains}, \eqref{eq: f2}, and \eqref{eq: f3} into  \eqqref{eq: f1}, we obtain:
\begin{equation}
\label{eq: f4}
\begin{aligned}
\delta P_{ext} &= \frac{1}{g^{3/2}} \ \iv{v}{}{,1} \cdot \left[ \left(\ve{m} \cdot \iv{g}{}{3}\right) \left( \iv{g}{}{2} \otimes \iv{g}{}{1} + \iv{g}{}{1} \otimes \iv{g}{}{2} \right) - \left(\ve{m} \cdot \iv{g}{}{2}\right) \left( \iv{g}{}{3} \otimes \iv{g}{}{1} + \iv{g}{}{1} \otimes \iv{g}{}{3} \right) \right. \\
& \left. - \left(\ve{m} \cdot \iv{g}{}{1}\right) \left( \iv{g}{}{3} \otimes \iv{g}{}{2} \right) \right] \delta \iv{v}{}{,1} + \frac{1}{g^{3/2}} \ \omega \left( \ve{m} \cdot \iv{g}{}{1} \right)\left( \iv{g}{}{1} \cdot \delta \iv{v}{}{,1} \right) - \frac{1}{\sqrt{g}} \ \omega \left( \ve{m} \cdot \delta \iv{v}{}{,1} \right).
\end{aligned}
\end{equation}
If we define the submatrices: 
\begin{equation}
\label{eq: f5}
\begin{aligned}
\iveq{G}{}{11} &= \frac{1}{g^{3/2}} \ \left[ \left(\ve{m} \cdot \iv{g}{}{3}\right) \left( \iv{g}{}{2} \otimes \iv{g}{}{1} + \iv{g}{}{1} \otimes \iv{g}{}{2} \right) - \left(\ve{m} \cdot \iv{g}{}{2}\right) \left( \iv{g}{}{3} \otimes \iv{g}{}{1} + \iv{g}{}{1} \otimes \iv{g}{}{3} \right) \right. \\
& \left. - \left(\ve{m} \cdot \iv{g}{}{1}\right) \left( \iv{g}{}{3} \otimes \iv{g}{}{2} \right) \right], \\
\iveq{G}{}{12} &= \iveq{G}{}{21} = \iveq{G}{}{22} = \textbf{0}_{3\times3}, \\
\iveq{G}{}{13} &= \iveq{G}{}{23} =\textbf{0}_{3\times1}, \\
\iveq{G}{}{31} &= \frac{1}{g^{3/2}} \  \left( \ve{m} \cdot \iv{g}{}{1} \right) \iv{g}{\text{T}}{1}  - \frac{1}{\sqrt{g}} \ \iv{m}{\text{T}}{}, \\
\iveq{G}{}{32} &= \textbf{0}_{1\times3}, \\
\iveq{G}{}{33} &= 0,
\end{aligned}
\end{equation}
of the matrix:
\begin{equation}
\label{eq: f6}
\iveq{G}{}{} = 
\begin{bmatrix}
\iveq{G}{}{11} & \iveq{G}{}{12} & \iveq{G}{}{13} \\
\iveq{G}{}{21} & \iveq{G}{}{22} & \iveq{G}{}{23} \\
\iveq{G}{}{31} & \iveq{G}{}{32} & \iveq{G}{}{33}
\end{bmatrix},
\end{equation}
then the matrix $\iveq{G}{}{}$ can be simply added to the matrix $\ve{G}$ in \eqqref{eq: apnestoG matrix def}.

\bibliography{SpatialCurvedBeamBorkovicetal} 
\bibliographystyle{ieeetr}

\end{document}